\documentclass[useAMS,usenatbib]{mn2e}

\usepackage{graphicx}
\usepackage{amssymb}
\usepackage{color}
\newcommand{\Teff}{\ensuremath{T_{\mathrm{eff}}}}
\newcommand{\logg}{\ensuremath{\log g}}

\newcommand{\bv}{\ensuremath{B\!-\!V}}
\newcommand{\uv}{\ensuremath{U\!-\!V}}
\newcommand{\vk}{\ensuremath{V\!-\!K}}
\newcommand{\ub}{\ensuremath{U\!-\!B}}
\definecolor{squares}{RGB}{255,48,0}
\definecolor{triangles}{RGB}{0,86,180}
\definecolor{crosses}{RGB}{0,158,115}
\defcitealias{Johnson2010}{J\&P10}
\defcitealias{vanLoon2007}{vL+07}
\defcitealias{Marino2012}{M+2012}

\title[Abundances and clustering analysis of $\omega$~Cen]{Spectral matching for abundances and clustering analysis of stars on the giant branches of $\omega$~Centauri}
\author[J. D. Simpson, P. L. Cottrell  and C. C. Worley]{Jeffrey. D. Simpson$^{1}$\thanks{E-mail:
jeffrey.simpson@pg.canterbury.ac.nz}, P. L. Cottrell$^{1}$ and C. C. Worley$^{2}$\\
$^{1}$Department of Physics and Astronomy, University of Canterbury, Private Bag 4800, Christchurch, New Zealand\\
$^{2}$Laboratoire Lagrange (UMR7293), Universit\'{e} de Nice Sophia Antipolis, CNRS,\\Observatoire de la C\^{o}te d'Azur, BP 4229, 06304, Nice Cedex 4, France}
\begin{document}

\date{\today}

\pagerange{\pageref{firstpage}--\pageref{lastpage}} \pubyear{2012}

\maketitle

\label{firstpage}

\begin{abstract}
We have determined stellar parameters and abundances for 221 giant branch stars in the globular cluster $\omega$~Centauri. A combination of photometry and lower-resolution spectroscopy was used to determine temperature, gravity, metallicity, [C/Fe], [N/Fe] and [Ba/Fe]. These abundances agree well with those found by previous researchers and expand the analysed sample of the cluster. $k$-means clustering analysis was used to group the stars into four homogeneous groups based upon these abundances.

These stars show the expected anticorrelation in [C/Fe] to [N/Fe]. We investigated the distribution of CN-weak/strong stars on the colour-magnitude diagram. Asymptotic giant branch stars, which were selected from their position on the colour-magnitude diagram, were almost all CN-weak. This is in contrast to the red giant branch where a large minority were CN-strong. The results were also compared with cluster formation and evolution models. Overall, this study shows that statistically significant elemental and evolutionary conclusions can be obtained from lower resolution spectroscopy.
\end{abstract}


\section{Introduction}
$\omega$~Centauri is a large star system in the southern sky. Its overall appearance is that of a globular cluster (GC) but it has several features that set it apart from the classical ``simple-stellar population'' globular cluster. Its main sequence is at least bifurcated \citep{Bedin2004,Piotto2005} and the giant branch (GB) shows multiple branches \citep[][and references therein]{Johnson2010} (herein referred to as J\&P10). These multiple pathways on the colour-magnitude diagram are chemically different in overall metallicity \citep{Norris1995} and helium abundances \citep{Dupree2011}. \citet{Marino2012} (herein referred to as M+2012) suggest that some of these difference could be explained by CNO variations.

A number of studies have examined the metallicity and abundance profile of $\omega$~Cen in detail. Cluster stars with metallicities from [Fe/H]$=-2.2$ to $-0.5$ have been observed spectroscopically. Most of the GB stars are found to be around [Fe/H]$=-1.75$ \citep{Norris1995,Stanford2006,Marino2011,Johnson2010}. There is a second peak in the metallicity distribution round $-1.50$, with evidence for further peaks at $-1.15$, $-1.05$ and $-0.75$ \citepalias{Johnson2010}. These peaks correspond to different GB that have been observed photometrically \citep{Rey2004}.

$\omega$~Cen differs from most mono-metallic clusters not only in its metallicity range but in its neutron-capture abundances. Most\footnote{One mono-metallic cluster with a neutron-capture spread is M15 \citep{Sobeck2011}.} mono-metallic clusters show no variation in their s- or r-process abundances \citep[e.g.\,in 47 Tuc as observed by][]{Worley2012}, while within $\omega$~Cen there is a positive correlation of s-process abundances with metallicity. In fact there is over a 1 dex increase in abundance of [La/Fe] and [Ba/Fe] between [Fe/H]$=-1.9$ and $-1.5$. Barium abundances from \citet{Villanova2010} have a 1.5 dex scatter at all [Fe/H]. This was also found for the light s-process elements, yttrium and zirconium. \citet{Villanova2010} reported evidence for bimodality in the [Ba/Fe] to [Fe/H] distribution. For stars with [Fe/H]$<-1.5$ there were two populations: [Ba/Fe]$\sim+1.0$ or $\sim-0.2$. This result does not agree with the larger samples of \citetalias{Johnson2010} and \citet{Marino2011} who found no such bimodality. Instead there was a continuous trend of increasing [Ba/Fe] with increasing [Fe/H], at least up to [Fe/H]$=-1.4$. Above this metallicity there is a flattening or even turning over of [Ba/Fe] with respect to [Fe/H] \citep{Stanford2010}.

There are a number of small studies of the carbon and nitrogen abundances in $\omega$~Cen. \citet{Brown1993} analysed six stars finding $-0.45<$[C/Fe]$<+0.1$ and $0.75<$[N/Fe]$<1.25$. These stars had a large range of metallicities but did not include any stars with [Fe/H]$>-1.25$. \citet{Norris1995} had a larger sample of 40 (biased) stars for which they determined a larger range than seen by \citet{Brown1993} for both carbon and nitrogen. They found that there was a ``floor'' to their [C/Fe] values at $\sim-0.8$, a feature that they pointed out was not due to their abundance analysis method. Their most carbon-rich objects were roughly solar. \citet{Stanford2010} examined 33 stars at low resolution again finding a large range of carbon and nitrogen values in the cluster. They reported an anticorrelation but no bifurcation in their results.

The most recent study and at the highest resolution ($\sim20,000$) is \citetalias{Marino2012}. They observed 77 giants and found a correlation between the overall ${\rm C}+{\rm N}+{\rm O}$ abundance and the metallicity. Using the sodium-oxygen anticorrelation, they were able to distinguish what they referred to as first and second generation stars. Their results show a clear bifurcation between CN-weak and CN-strong stars (their Figure 2). The nitrogen-rich stars are grouped together with little spread in carbon. The CN-weak stars show a range of [C/Fe] from $-0.5$ to $+0.5$. With 221 stars, the work presented in our study will expand upon these stars as there are only 10 stars in common with \citetalias{Marino2012}.

One of the important ingredients in any theory of the formation of $\omega$~Cen is the age spread of the stars. A variety of ranges have been reported in the literature. The shortest age differences have been reported by \citet{Sollima2005} who found 2 Gyr. At the opposite end is \citet{Villanova2007} who found up to 5 Gyr. The placement of the isochrones that are used in this sort of age determination depends on the CNO abundance of the models. \citetalias{Marino2012} used the \citet{Pietrinferni2006} isochrones and found that ``for a fixed turn-off and sub-giant branch (SGB) brightness, CNO-enhanced isochrones provide younger ages than isochrones corresponding to a canonical $\alpha$-enhanced mixture; as a rule-of-thumb we found the $\delta{\rm age}/\delta{\rm [CNO]}\sim-3.3$Gyr per dex''.

Multi-object spectrographs allow for hundreds of stars to be observed in one night, making it possible to efficiently sample an object like a globular cluster with $10^4-10^6$ stars. Recent papers such as \citet{Marino2011} used the VLT to obtain spectra of 300 stars and \citetalias{Johnson2010} published results for 855 stars. They confirmed the peculiar chemical make-up of $\omega$~Cen. Our study makes use of both of these datasets and extends them with an investigation of the \citet{vanLoon2007} (herein referred to as vL+07) spectral library. The work presented here determined the common stellar parameters (temperature, surface gravity, metallicity) as well as s-process, carbon and nitrogen abundances using lower-resolution spectra and photometric data of this sample of $\omega$~Cen giant branch stars.

Section \ref{sec:Observations} discusses the observational data sources; Section \ref{sec:DataReduction} explains the methods for analysing the \citetalias{vanLoon2007} spectral library; Section \ref{sec:Results} presents the results and quality assurance; Section \ref{sec:ClusteringAnalysis} presents a statistical clustering analysis of the dataset from this work and of \citet{Marino2011,Marino2012}; Section \ref{sec:AGB} shows that the easily identifiable asymptotic giant branch (AGB) stars are likely to be CN-weak; Section \ref{sec:EvolutionaryModels} discusses the results in the context of evolutionary models proposed in previous works.

\section{Observational data sources}\label{sec:Observations}

\begin{table}
\caption{Overview of different data sources used in this work.}
\label{Table:DataSources}
\begin{tabular}{llr}
\hline
Source & Type & Stars\\
\hline
\citetalias{vanLoon2007} & Spectroscopy & 1518 \\
\citetalias{Johnson2010} & Abundances & 855\\
\citet{Bellini2009a} & Photometry & 359,391\\
{\sc 2mass} & Photometry & 26,985 \\
\citetalias{Marino2012} & Abundances & 77\\
\hline
\end{tabular}
\end{table}

A dataset of 221 giant branch stars (including 14 stars that were observed twice spectroscopically) was created by combining information from \citetalias{vanLoon2007}, {\sc 2mass} \citep{Skrutskie2006}, \citet{Bellini2009a} and \citetalias{Johnson2010}. Table \ref{Table:DataSources} describes the data sources that were combined and used in this work.

The datasets, and how they were combined are described in this section. Section \ref{sec:normalization} describes the \citetalias{vanLoon2007} spectral library, Section \ref{sec:photometry} describes the photometric data sources, with the positional matching of different datasets described in Section \ref{sec:positional_matching}. The final selection of stars is found in Section \ref{sec:selection}.

\subsection{\protect\citetalias{vanLoon2007} spectral library}\label{sec:normalization}
The \citetalias{vanLoon2007} dataset is a spectral library of 1518 post-main sequence stars in $\omega$~Cen. Spectra were obtained with the 2dF instrument at the Anglo-Australian Telescope, covering approximately $\lambda\sim3840$--4940 \AA\ at a resolving power of $\lambda/\Delta\lambda\sim1600$ and a signal-to-noise per pixel ranging from $\sim50$ in the blue to $>100$ in the redder part of the wavelength range. For each star \citetalias{vanLoon2007} provided a unique ID from \citet{vanLeeuwen2000} (known as the LEID), positional information, a $B$ magnitude, a \bv\ colour and a radial velocity. \citetalias{vanLoon2007} also did a stellar parameter determination for their catalogue. This was done solely from the spectra and assumed solar-scaled abundances. For our study, we decided to use their raw spectra and analyse the stars ourselves with appropriate globular cluster abundances and concentrating on small sections of the spectrum.


We iteratively continuum normalized the raw spectrum by mapping the spectrum onto a synthetic spectrum of the best stellar parameters. Three anchor points were selected: 4088\AA, 4220\AA\ and 4318\AA\ \citep[as also used in][]{Worley2012}. These three points were joined by two straight lines and the intensities were mapped from the raw spectrum onto the synthetic spectrum. This method is dependent on the synthetic spectrum being used and the continuum placement was recomputed at each iterative step using the method described in Section \ref{sec:SpectralMatching}.

At the resolution of the \citetalias{vanLoon2007} spectra, and with its wavelength coverage, it was not possible to use a spectral-temperature indicator such as H$\beta$. Therefore the temperature-colour relationships of \citet{Alonso1999} were used. This could then be combined with the luminosity relationship to estimate the surface gravity of the stars. For stars of unknown metallicities, the most appropriate colour to use is \vk, as noted by \citet{Alonso1999} in their section 2.3. Consequently, additional photometry of the stars in the \citetalias{vanLoon2007} catalogue was needed, since they only had $B$ and $V$.

\subsection{Photometry data sources}\label{sec:photometry}

\subsubsection{{\sc 2mass} photometry}
The Two Micron All Sky Survey ({\sc 2mass}) \citep{Skrutskie2006} observed the sky in the near-infrared covering three photometric bands: $J$ (1.25 $\mu$m), $H$ (1.65 $\mu$m), and $K_s$ (2.17 $\mu$m). The {\sc 2mass} data were sampled for a circular region with a radius of 28\arcmin\arcmin\ around RA~$=13^\mathrm{h}26^\mathrm{m}14.16^\mathrm{s}$, ${\rm Dec}=-47^\circ31\arcmin11.95\arcmin\arcmin$, the midpoint of the \citetalias{vanLoon2007} data. This created a catalogue of 26,985 stars with $J$, $H$ and $K_S$ magnitudes in the {\sc 2mass} photometric system. The majority of the stars in the raw dataset were from the globular cluster with obvious contamination by field stars. The photometry extended well below the cluster's main sequence turn off, covering all the \citetalias{vanLoon2007} stars. 

\subsubsection{\citet{Bellini2009a} photometry}
\citet{Bellini2009a} used ESO archive data to create a new, CCD-based, proper-motion catalogue for $\omega$~Cen, extending to $B\sim20$, containing 359,391 stars with $U$, $B$, $V$, $R_{\rm C}$, $I_{\rm C}$, and H$\alpha$. They used 279 archive images acquired at the ESO/MPI2.2m telescope at La Silla (Chile) equipped with the wide-field imager camera (WFI).  It covered 83\% of the stars of the \citetalias{vanLoon2007} dataset. This catalogue improved upon the previous large proper motion catalogue of $\omega$~Cen, produced by \citet{vanLeeuwen2000} from photographic plates. Although \citetalias{vanLoon2007} had $B$ and $V$ magnitudes for all of their stars, it was decided to use the \citet{Bellini2009a} photometry instead as it provided a very precise colour-magnitude diagram that could be used to distinguish the metal-poor AGB (the bluest stars on the GBs)from all of the red giant branch (RGB). Other AGBs will be mixed with the RGBs as a consequence of the different evolutionary paths followed by stars of different ages and metallicities.

\subsection{Positional matching} \label{sec:positional_matching}
To determine the \Teff\ and \logg\ of the stars, the \vk\ colour was needed. This required the positional matching of \citetalias{vanLoon2007} sources and the photometric sources of  \citet{Bellini2009a} and {\sc 2mass}. The closest positional match was found. 

{\sc 2mass} stated that their positional accuracy should be less than 130 mas for saturated sources and $<80$ mas for the best unsaturated sources. The \citet{vanLeeuwen2000} dataset, from which the \citetalias{vanLoon2007} is derived, had positional errors mostly less than 100 mas in right ascension and declination. \citet{Bellini2009a} stated that based upon $\sim5500$ reference stars, their error was 45--50 mas in each coordinate. The positional matching of the three photometric catalogues was done using {\sc topcat} \citep{Taylor2005}. 982 stars were found to be in common between \citetalias{vanLoon2007} and {\sc 2mass}, with 672 of these stars also matching to \citet{Bellini2009a}.

\subsection{Selection of stars}\label{sec:selection}

\begin{table}
\caption{Photometry of the 221 stars. The LEID is the star label given by \citetalias{vanLoon2007}. The $U$ and $V$ photometry is from \citet{Bellini2009a} and the $J$ and $K$ photometry is from {\sc 2mass}. All the photometry is in their original system. The number of spectra of each star in the \citetalias{vanLoon2007} spectral library is given.}
\label{Table:Photometry}
\begin{tabular}{rrrrrc}
\hline
LEID & U$^{a}$  & V$^{a}$ & J$^{b}$ & K$^{b}$ & N. of Spectra \\
\hline
15022 & 14.19 & 12.10 & 9.730 & 8.95 & 1\\
15023 & 14.15 & 12.26 & 9.964 & 9.23 & 1\\
16019 & 14.78 & 13.28 & 11.268 & 10.62 & 2\\
16027 & 14.83 & 13.55 & 11.532 & 10.86 & 1\\
17029 & 14.42 & 13.17 & 11.201 & 10.54 & 1\\
17046 & 14.41 & 13.31 & 11.364 & 10.74 & 1\\
18040 & 15.10 & 12.79 & 10.448 & 9.60 & 2\\
19022 & 14.30 & 13.16 & 11.212 & 10.58 & 2\\
19062 &  & 12.85 & 10.601 & 9.87 & 1\\
\hline
\end{tabular}
(This table is available in its entirety in a machine-readable form in the online journal. A portion is shown here for guidance regarding
its form and content.)
\end{table}

\begin{figure}
\includegraphics[width=84mm]{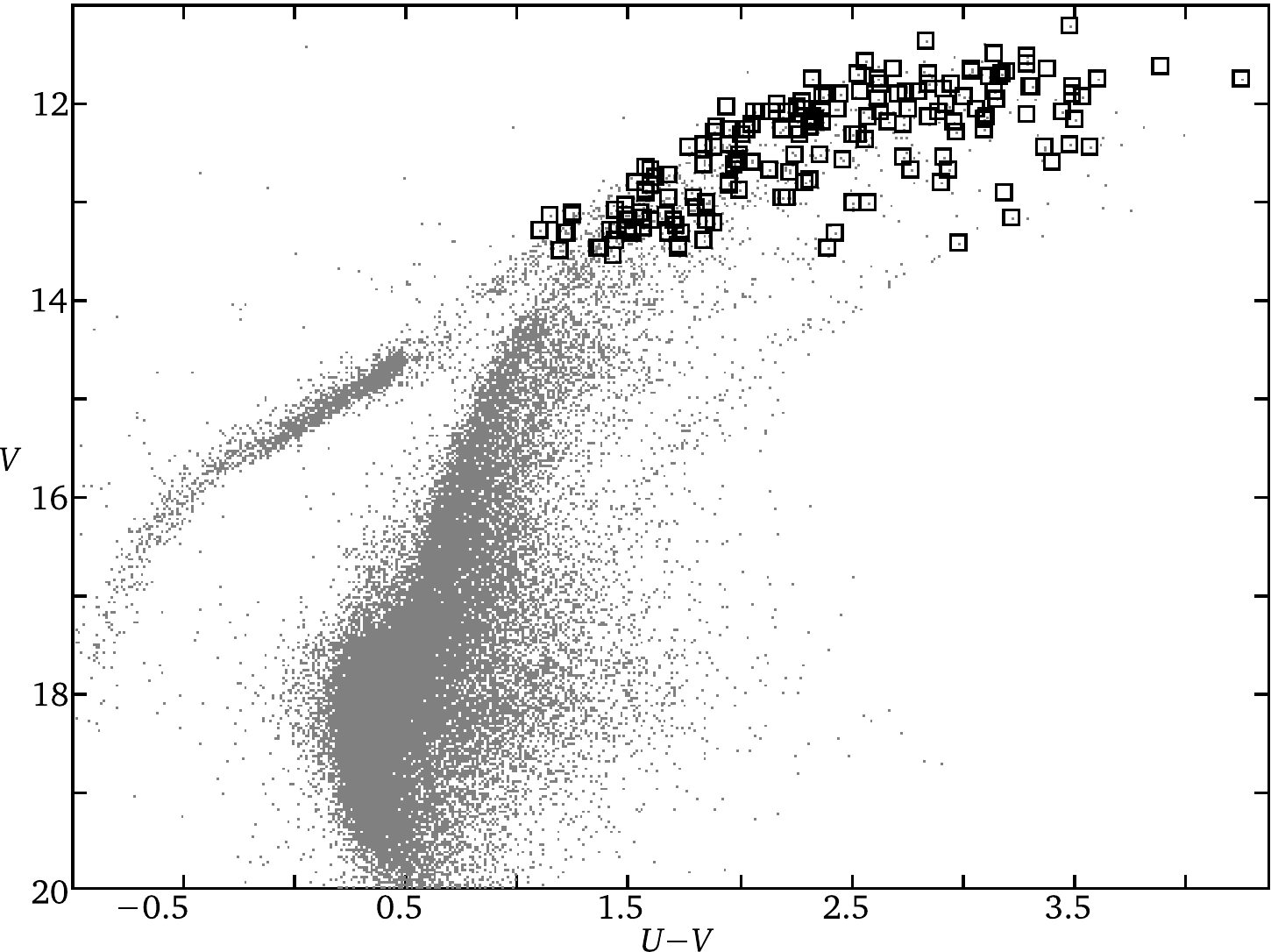}
\caption{A $V$-(\uv) colour-magnitude diagram of $\omega$~Cen derived from the \citet{Bellini2009a} photometry (small dots). The open squares highlight the positions of 172 (of 211; the missing stars did not have $U$ photometry) stars in our sample. The magntiude limit of $V\sim13.5$ is due to the limiting magnitude of \citetalias{Johnson2010}. The abundances determined are presented in Table \ref{Table:Johnson2010}. \label{fig:cmd_all_stars}}
\end{figure}

\citetalias{Johnson2010} determined several elemental abundances including [Fe/H], [O/Fe], [Na/Fe] and [Ca/Fe]. For the analysis presented in this work it was decided to concentrate on stars that were also in the \citetalias{Johnson2010} catalogue as it would result in more precise abundances that did not have further uncertainty due to unknown oxygen and calcium abundances. A full analysis of the complete \citetalias{vanLoon2007} spectral library, with reasonable estimates of oxygen and calcium, is in preparation.

\citetalias{Johnson2010} selected their stars from \citet{vanLeeuwen2000} and there were 291 matches in total between \citetalias{Johnson2010} and \citetalias{vanLoon2007}. Not all of these stars were positionally matched to the \citet{Bellini2009a} and {\sc 2mass} libraries. Limiting to just those stars that were, a final catalogue of 221 stars was created (Figure \ref{fig:cmd_all_stars} and Table \ref{Table:Photometry}).

This dataset does have a selection bias against the core of the cluster. This is due to the crowding that makes it difficult for a best positional match to identify the correct star. Comparing the radial distributions of the \citetalias{Johnson2010} dataset and our own, we find that their stars are much more centrally concentrated: 57\% of the stars in the \citetalias{Johnson2010} dataset are within $6\arcmin$ of the cluster centre while for our set it is only 29\%. In Section \ref{sec:CN} we note that this selection effect does not appear to have greatly biased our sample, as we have a similar proportion of O-rich and O-poor stars as found by \citetalias{Johnson2010}.

Overall, our 221 stars have a magnitude range of $11.2<V<13.6$ and a colour range of $0.8<(\bv)<1.8$. This corresponds to stars with \Teff\ from 3800~K to 4900~K and \logg\ from 0.3 to 1.8 (Section \ref{sec:temp_grav}).

\section{Data reduction of spectral library}\label{sec:DataReduction}

\subsection{Temperature and gravity}\label{sec:temp_grav}
The temperature-colour relationship of \citet{Alonso1999} was used to calculate the \Teff\ using their equations 8 \& 9. These equations have constraints on \vk\ and [Fe/H]. If the star satisfied the constraints for both equations then an average of the two temperatures was used. If neither set of constraints were satisfied then the star was excluded.

\citet{Alonso1999} required the \vk\ photometry to be in the Carlos S\'{a}nchez Telescope photometric system. The photometry of \citet{Bellini2009a} and {\sc 2mass} photometry was converted to this system using a method developed by \citet{Johnson2005}. The interstellar reddening adopted was $E(\vk)=0.3$ \citep{Johnson2008}. \citet{Calamida2005} found evidence for an intra-cluster reddening variation in $\omega$~Cen of $0.06\leq{\rm E}(\bv)\leq0.13$. In terms of the effect on the \Teff, this would create a range in \Teff\ of $\sim70$K for the hottest stars. Such a change in temperature is of a comparable order of magnitude to the uncertainties resulting from the photometric errors.

Differential reddening was disputed by \citet{Villanova2007}. They produced CMDs of a million stars from ACS photometry. They point to the sharpness of their CMDs as evidence for the lack of any range of intra-cluster reddening for $\omega$~Cen. So although the range of reddening suggested by \citet{Calamida2005} would have a measurable effect on the temperatures, it did not seem to be a significant issue across the whole cluster.

The surface gravity was determined using the elementary equation,
\begin{eqnarray*}
\logg_* &=& 0.40(M_\mathrm{bol,*}-M_{\rm bol,{\rm\sun}}+\logg_{\rm\sun}\\
&&+ 4\log (T_*/T_{\rm\sun})+\log (M_*/M_{\rm\sun}),
\end{eqnarray*}
with $M_{\rm bol,{\rm\sun}}=+4.75$, $T_{\rm\sun}=5777$~K and $\logg_{\rm\sun}=4.44$ \citep{Allen2000}. The stellar mass was set to 0.8~$M_{\rm\sun}$. There is evidence for an age range in $\omega$~Cen which would imply a mass range. This mass range would extend down to 0.6~$M_{\rm\sun}$. However, since the logarithm of the stellar mass is used here, this reduced the effect on the final gravity value to 0.1 dex at the most extreme. As such, the probable mass range was ignored. The bolometric correction was taken from \citet{Alonso1999} and there is a small [Fe/H] dependence which, across the metallicity range of $\omega$~Cen would affect the surface gravity by 0.01 dex.


Differences in gravity can be mimicked by differences in metallicity, especially with the ionized barium line at 4554\AA\, used in this study for determining [Ba/Fe]. Increasing the gravity of a star will decrease the strength of spectral lines of ionized species (for all other parameters remaining constant). It was found that a 0.3~dex change in gravity (for a $\Teff=4250$K, $\logg=1.6$ and [Fe/H]$=-0.5$ star) could be compensated for by a 0.05~dex change in the metallicity. For a gross change of metallicity it was not possible to use the gravity to compensate.

\citetalias{Johnson2010} and \citet{Marino2011,Marino2012} all found essentially the same temperatures and gravities for our sample of stars as they used the same method \citep{Johnson2005} and photometry \citep{vanLeeuwen2000,Bellini2009a}. Any small differences were due to distance moduli, reddening and stellar mass selections.

\subsection{Spectral matching}\label{sec:SpectralMatching}
\begin{figure}
\includegraphics[width=84mm]{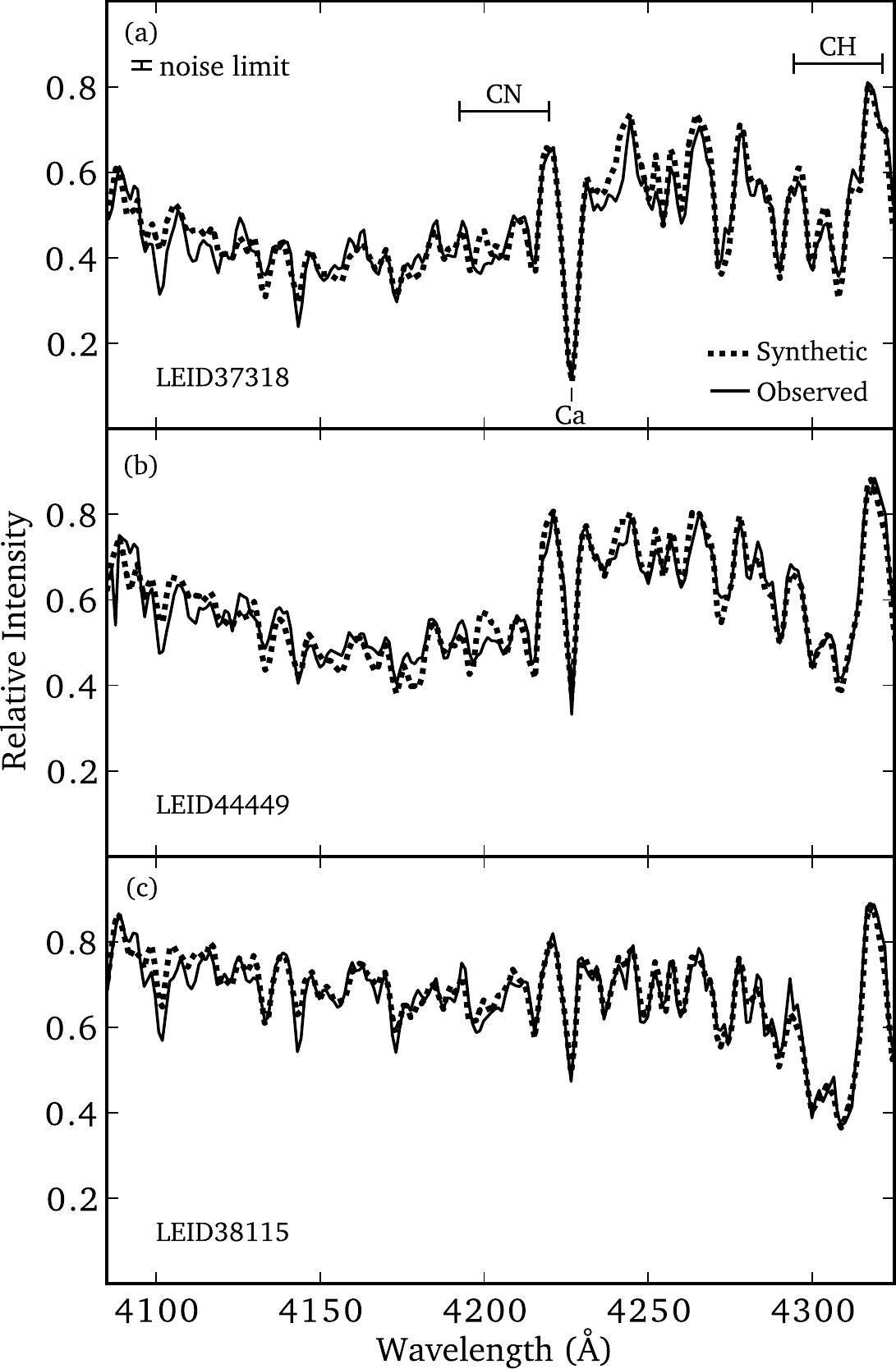}
\caption{Examples of three spectra. (a) is an RGB-a star with a high metallicity ([Fe/H]$=-1.0$), illustrating the three regions that were used in to determine the metallicity (via the Ca 4224\AA\ line), carbon (CH) and nitrogen (CN) abundances. The noise level is shown to reflect the signal-to-noise in the spectra. (b) and (c) are a CN-weak/-strong pair and show the variation in the CN band. The difference in [N/Fe] between (b) and (c) is 1.4 dex. \label{fig:spectra_examples}}
\end{figure}
Spectrum synthesis was performed using {\sc moog} \citep{Sneden1973}. {\sc moog} is a code that performs a variety of {\sc lte} line analysis and spectrum synthesis tasks. Around this was created a Python wrapper to find the best matching synthetic spectrum using known parameters and physically reasonable estimates for unknown parameters of the stars.

For each star, the temperature and gravity were calculated from the \vk\ photometry and assumptions about the stellar mass as described in Section \ref{sec:temp_grav}. The metallicity was also known for all of these stars from \citetalias{Johnson2010}. To keep the results of [Fe/H], [C/Fe] and [N/Fe] self-consistent, [Fe/H] was determined for all the stars using our spectral matching method. Being able to determine [Fe/H] will be required when working with the full \citetalias{vanLoon2007} dataset as the metallicities are not known for the majority of these stars.

At the resolution of the \citetalias{vanLoon2007} spectra, there was not enough information to try and determine an [O/Fe] abundance, so the \citetalias{Johnson2010} values were used in this work. The other known value for all of the stars was [Ca/Fe] which were also taken from \citetalias{Johnson2010}.

The \Teff, \logg\ and [Fe/H] values were used to generate the necessary model atmospheres via interpolation\footnote{The interpolation routine used atmosphy https://github.com/andycasey/atmosphy.} within the $\alpha$-enhanced grid of \citet{Castelli2004}. A value of [$\alpha$/Fe]$=0.3$ was used \citepalias{Johnson2010}. For the CH and CN region of the spectrum, line lists were from \citet{Norris2012}. These line lists were found to match to the high resolution atlas of Arcturus \citep{Hinkle2000} using abundances determined by \citet{Decin2004}. They also matched the Arcturus spectrum convolved to the resolution of the \citetalias{vanLoon2007} spectra. The atomic $\log gf$ values in the line list for the barium region investigated (400 to 500 nm) were adjusted so that the synthetic spectrum matched the Arcturus spectrum, again at both resolutions.

Lines of barium exhibit hyperfine splitting (hfs). In spectrum synthesis, hfs will cause the line to have a larger equivalent width for a given abundance compared to when hfs is not included. For the 4554\AA\ line of barium, two line lists were created: one that included all the isotopic species of the barium feature \citep{Rutten1978} and one that did not. It was found that across the temperature and gravity ranges of the cluster, the effect was negligible. Such results confirmed what was found by \citet{Stanford2010} for main sequence stars. They found a difference of [Ba/Fe]$<0.05$ dex between abundances found with hfs and those without.

The microturbulence ($v_t$) was assumed to be 2.0 km\,s$^{-1}$ for all of the stars. This is within the range of values determined by \citetalias{Johnson2010}. We were unable to refine it further due to the lack of unblended iron lines to remove the trend in iron abundance with line strength.

The code we developed read in the spectrum of the star along with previously determined stellar parameters (\Teff, \logg, [O/Fe] and [Ca/Fe]) for the star and then found the best fitting synthetic spectrum created using {\sc moog}. It was designed to search the parameter space in a systematic way. The S(3839) index from \citetalias{vanLoon2007} was used as an estimate as to whether the star is CN-weak and CN-strong. This gave the initial guesses for [C/Fe] and [N/Fe]. An iterative process was used. At each step of the process a model atmosphere and then synthetic spectrum was created from the full parameter set (\Teff, \logg, [Fe/H], [O,Ca,C,N,$\alpha$/Fe]). The [Fe/H] was determined using the metallicity sensitive CaI line at 4226\AA. The other two spectral features used were the carbon-dependent G band (mainly CH: $A^2\Delta-X^2\Pi$) region (4295\AA\ to 4325\AA) and the CN blue-system ($B^2\Sigma-X^2\Sigma$) (4195\AA\ to 4222\AA).

At the first step, the initial guess for [C/Fe] and [N/Fe] was used to find [Fe/H] using the CaI and the known [Ca/Fe] ratio \citepalias{Johnson2010}, minimizing the difference in the equivalent width between the synthetic and observed spectra for the calcium line. Then the newly determined [Fe/H] was used in the $\chi^2$ minimization to find [C/Fe] from the G band. Finally the determined [Fe/H] and [C/Fe] were used to determine [N/Fe] from the CN-region. This process was then repeated, with the newly determined parameters from the previous iteration. [Fe/H] was redetermined using the new [C/Fe] and [N/Fe] values, requiring the raw spectrum to be continuum normalized to the new synthetic spectrum. This full cycle was repeated until it converged on an unchanging triplet of [Fe/H], [C/Fe] and [N/Fe]. A sample of three spectra is shown in Figure \ref{fig:spectra_examples}.

\begin{table*}
\begin{minipage}{137mm}
\caption{Results of the delta analysis performed on four stars to understand the effect of changing the input temperature by $+100$~K and the gravity by $+0.3$~dex.}
\label{Table:DeltaAnalysis}
\begin{tabular}{rrrrrrrrr}
\hline
LEID & \Teff & \multicolumn{2}{c}{[Fe/H]} & \multicolumn{2}{c}{[C/Fe]} & \multicolumn{2}{c}{[N/Fe]} \\
&&$\Delta\Teff=+100$~K&$\Delta\logg=+0.3$ dex&+100~K&+0.3 dex&+100~K&+0.3 dex\\
25062&3965&$-0.1$&$+0.1$&$-0.3$&$-0.2$&$+0.1$&$+0.3$\\
61085&3965&$+0.2$&0.0& $-0.2$& 0.0& $-0.4$& $+0.1$ \\
30013&4485&$-0.1$&0.0& 0.0& 0.0& $-0.1$& $+0.1$ \\
57114&4470&$+0.2$&$-0.1$& 0.0& 0.0& $-0.1$& $+0.1$ \\
\hline
\end{tabular}
\end{minipage}
\end{table*}
A delta analysis was performed to understand the sensitivity of the spectral matching method to the input parameters of temperature and gravity. Four stars were selected, one from each group identified in Section \ref{sec:ClusteringAnalysis}, with two stars being low temperature ($\sim4000$~K) and two being higher ($\sim4500$~K). The results of this analysis are presented in Table \ref{Table:DeltaAnalysis}. It was found that increasing the temperature by 100~K results in 0.1--0.2 dex increase in [Fe/H] and [C/Fe]. [N/Fe] changed by $\pm0.1$ dex except for the most N-rich star which decreased by 0.4 dex. Changing the gravity by +0.3 dex had little to no effect on the abundances (see also comments in Section \ref{sec:temp_grav}).

The first 20 stars (which have a total of 25 spectra) were analyzed by repeating their analysis with four random starting positions. This was to test the convergence of the method. All but three of the stars returned the values affected by less than 0.1 dex. The other three stars were all very N-poor stars and were among a group of stars that were excluded from the comparative analysis. At the spectral resolution of the \citetalias{vanLoon2007} library, there is not a lot of sensitivity to nitrogen at low values. From this, we would state that the convergence of our method is excellent for stars with [N/Fe]$>-0.5$.

\section{Results}\label{sec:Results}
\begin{table*}
\begin{minipage}{160mm}
\caption{Stellar parameters determined for the 221 stars. The LEID is the star label given by \citetalias{vanLoon2007}. Abundances from other studies are indicated. The offsets described in Section \ref{sec:Results} and Figure \ref{fig:CFe_M12_+03} have been applied to [C/Fe] and [N/Fe]. The group number is described in Section \ref{sec:ClusteringAnalysis}.}
\label{Table:Johnson2010}
\begin{tabular}{rrrrrrrrrrrr}
\hline
LEID & \Teff & \logg & [Fe/H] & [C/Fe] & [N/Fe] & [O/Fe]$^{a}$ & [Na/Fe]$^{a}$ & [Ba/Fe] & S(3839)$^{b}$ & CH(4300)$^{b}$ & Group \\
\hline
15022 & 4400 & 0.9 & $-1.8$ & $-0.5$ & 0.3 & 0.44 & $-0.36$ & 0.4 & 0.108 & 0.229 & 2\\
15023 & 4450 & 1.0 & $-1.6$ & $-0.6$ & 0.1 & 0.42 & $-0.29$ & $-0.2$ & 0.075 & 0.264 & 1\\
16019 & 4800 & 1.6 & $-1.7$ & 0.7 & $-0.1$ & 0.64 & 0.01 & 0.8 & 0.475 & 0.400 & 2\\
16027 & 4750 & 1.7 & $-2.3$ & $-0.1$ & $-1.0$ & 0.46 & $-0.16$ & $-0.1$ & 0.089 & 0.239 & 1\\
17029 & 4800 & 1.6 & $-1.9$ & $-0.8$ & $-0.9$ & 0.35 & 0.49 & $-0.2$ & 0.090 & 0.135 & 1\\
17046 & 4850 & 1.7 & $-2.3$ & $-0.6$ & 1.0 & 0.14 & 0.16 & 1.0 & 0.049 & 0.120 & 3\\
37318 & 3950 & 0.8 & $-1.0$ & $-0.4$ & 2.0 & 0.35 & 1.03 & 0.5 & 0.411 & 0.337 & 4\\
38115 & 4150 & 0.8 & $-1.6$ & 0.4 & 0.1 & 0.28 & 0.14 & 0.2 & 0.397 & 0.422 & 2\\
44449 & 4100 & 0.6 & $-1.5$ & $-0.6$ & 1.5 & $-0.73$ & 0.47 & 0.5 & 0.572 & 0.409 & 3\\
\hline
\end{tabular}

$^{a}$ \citetalias{Johnson2010}

$^{b}$ \citetalias{vanLoon2007}

(This table is available in its entirety in a machine-readable form in the online journal. A portion is shown here for guidance regarding
its form and content.)
\end{minipage}
\end{table*}

\begin{table*}
\begin{minipage}{160mm}
 \caption{Stellar parameters determined (this work $\equiv$ SCW) for the nine stars in common with \citetalias{Marino2012}. The LEID is the star label given by \citetalias{vanLoon2007} and the ID is the numbering scheme from \citetalias{Marino2012}. The [Fe/H], [C/Fe], [N/Fe] from \citetalias{Marino2012} are shown for comparison.}
 \label{Table:Marino2012}
 \begin{tabular}{rcrrrrrrrrrr}
 \hline
 LEID & ID & &  & \multicolumn{2}{c}{[Fe/H]} & \multicolumn{2}{c}{[C/Fe]} & \multicolumn{2}{c}{[N/Fe]} & \multicolumn{2}{c}{[O/Fe]} \\
 \citetalias{vanLoon2007}     & \citetalias{Marino2012}    & \Teff      &  \logg     & SCW & \citetalias{Marino2012} & SCW & \citetalias{Marino2012} & SCW & \citetalias{Marino2012} & \citetalias{Johnson2010} & \citetalias{Marino2012} \\
 \hline
35090 & 246585 & 4100 & 0.8 & $-1.3$ & $-1.64$ & $-0.55$ & $-0.65$ & 1.05 & 1.55 &$-0.32$ & $-0.25$\\
36156 & 245724 & 4150 & 0.7 & $-2.0$ & $-1.97$ & $-0.45$ & $-0.44$ & $-0.05$ & 0.20& 0.43 & 0.30\\
36179 & 244812 & 4050 & 0.8 & $-1.5$ & $-1.40$ & $-0.75$ & $-0.17$ & 2.05 & 1.40& $-0.42$ & $-0.10$\\
37253 & 242745 & 4150 & 0.7 & $-1.9$ & $-1.91$ & $-0.25$ & $-0.35$ & 0.15 & 0.10& 0.43 &0.37\\
38115 & 241359 & 4200 & 0.8 & $-1.6$ & $-1.69$ & $ 0.35$ &   0.17  & 0.05 & 0.26& 0.28 &0.40\\
46150 & 224500 & 3950 & 0.6 & $-1.6$ & $-1.51$ &   0.25  &   0.40  & 0.65 & 0.20& 0.44 &0.49\\
46194 & 225246 & 4200 & 0.9 & $-1.9$ & $-1.87$ & $-0.65$ & $-0.61$ & 0.65 & 0.70& $-0.05$ &0.19\\
48235 & 220325 & 4050 & 0.7 & $-1.5$ & $-1.61$ & $-0.65$ & $-0.45$ & 1.15 & 1.20& $-0.32$ &0.00\\
51091 & 215367 & 4400 & 1.1 & $-2.0$ & $-1.93$ & $-0.55$ & $-0.98$ & 0.75 & 0.90& 0.28 &0.15\\
\hline
 \end{tabular}
\end{minipage}
\end{table*}

\begin{figure}
\includegraphics[width=84mm]{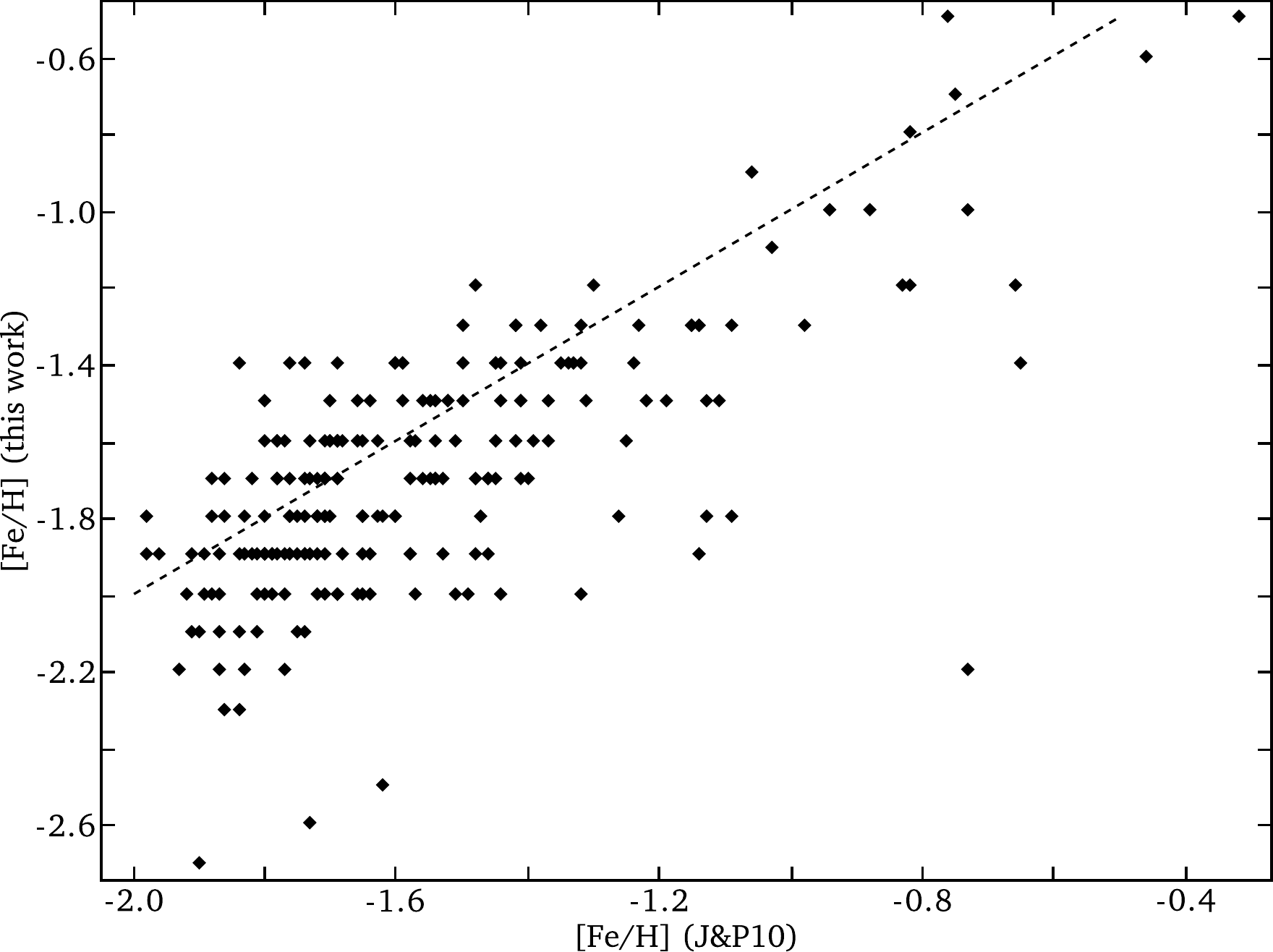}
\caption{Direct comparison of the metallicities found for each star in this study and the abundances found for the same stars by \citetalias{Johnson2010} (with the one-to-one line shown) \label{fig:FeH_J10_S12}}
\end{figure}

For each star the following parameters were determined: \Teff, \logg, [Fe/H], [C/Fe], [N/Fe] and [Ba/Fe] (Table \ref{Table:Johnson2010}). The CMD position of the stars is shown in Figure \ref{fig:cmd_all_stars}. There were 221 stars that were analysed with 14 having two spectra. An estimate of the uncertainty in the abundances determined was made using the doubly observed stars. The average standard deviation in the difference in [Fe/H] was $\pm0.2$~dex, [C/Fe] was $\pm0.2$~dex, [N/Fe] was $\pm0.3$~dex and [Ba/Fe] was $\pm0.6$~dex. Among the 221 stars previously mentioned there were nine stars in common between \citetalias{Marino2012} and \citetalias{vanLoon2007} (Table \ref{Table:Marino2012}). These stars were used for direct comparison of the [C/Fe] and [N/Fe] abundances as \citetalias{Johnson2010} did not determine these elements.

The abundances determined were compared to \citetalias{Johnson2010} and \citetalias{Marino2012} directly and indirectly. Plotting the metallicity found by this work against that of \citetalias{Johnson2010} shows a good correlation (Figure \ref{fig:FeH_J10_S12}). Most of the stars had [Fe/H] from $-2.0$ to $-1.5$, resulting in a large concentration around these values. As would be expected when results derived from spectra with a resolution of 1,600 is compared to spectra with a resolution of 20,000, there is a spread of values. About 70\% of the stars were within $\pm0.2$~dex of the value found by \citetalias{Johnson2010}. The average difference for all the stars was +0.1 dex: \citetalias{Johnson2010} found a slightly more metal-rich value than that found by this work.

The nine stars in common between this study and \citetalias{Marino2012} (Table \ref{Table:Marino2012}) had a range of \Teff\ from 4050 to 4400~K and in metallicity from $-1>{\rm [Fe/H]}>-2$. According to the abundances determined by \citetalias{Marino2012}, four would be classed as CN-strong ([N/Fe]$>0.6$), while four were CN-weak with higher carbon abundances. This meant that they provided a good sample across the range of stellar parameters which was not clumped at a particular metallicity or [X/Fe].

\begin{figure}
\includegraphics[width=84mm]{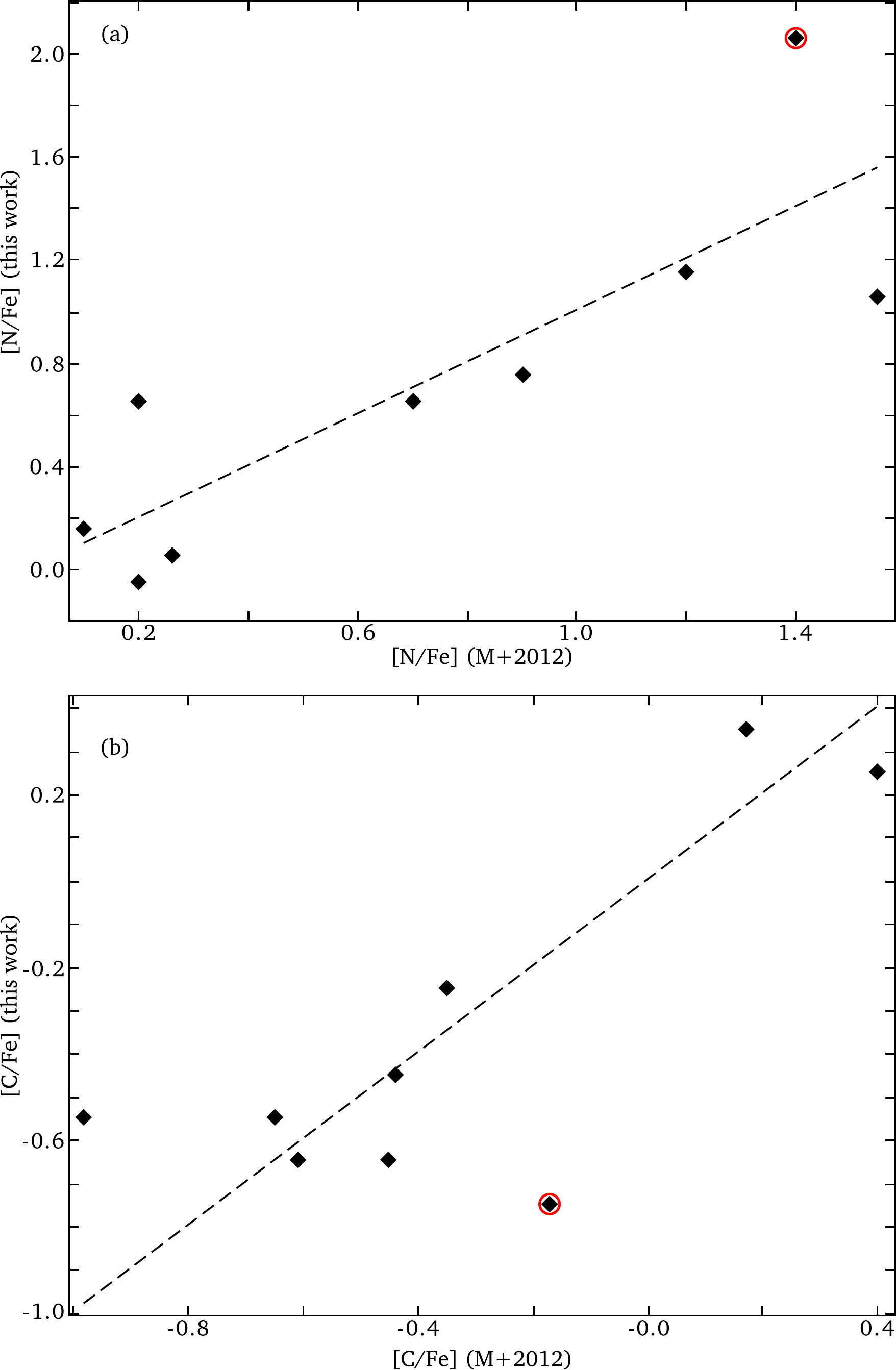}
\caption{(a) [N/Fe] and (b) [C/Fe] abundances found by this work and \citetalias{Marino2012}. The dashed line is the one-to-one line. An offset of $-0.55$ dex has been applied to our initial [N/Fe] abundances and $+0.45$ dex applied to our initial [C/Fe] abundances. This was done to place our results on the same abundance system as \citetalias{Marino2012}. The circled star is LEID36179, the outlier discussed in Section \ref{sec:Results}. \label{fig:CFe_M12_+03}}
\end{figure}

\begin{figure}
\includegraphics[width=84mm]{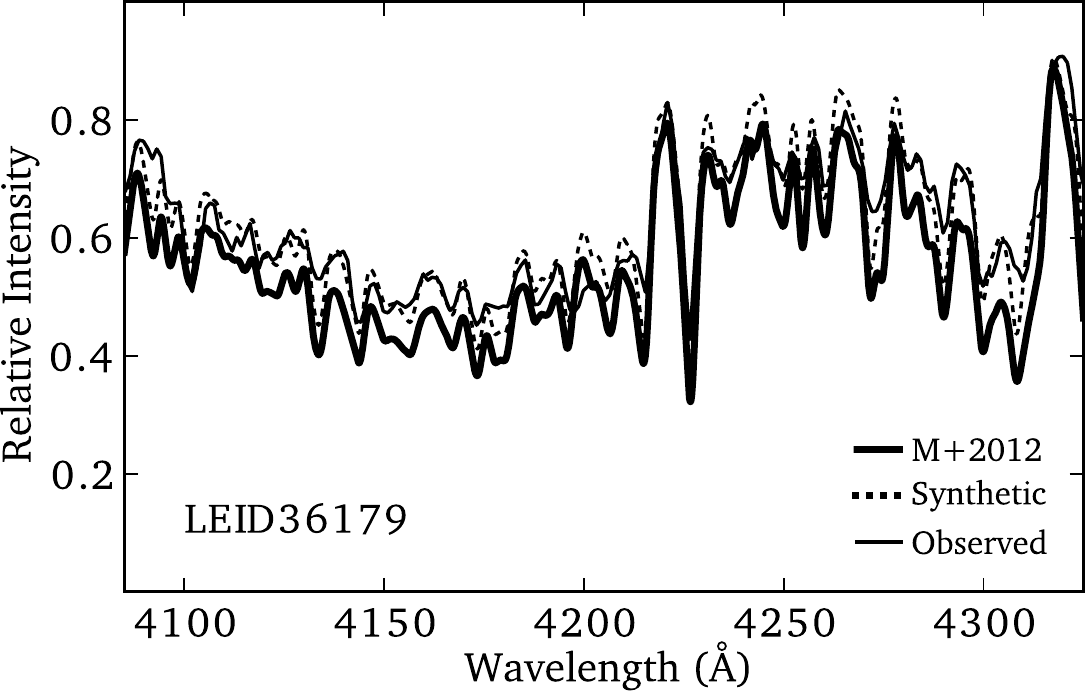}
\caption{LEID36179 is the significant outlier on figure \ref{fig:CFe_M12_+03}a. We determined [N/Fe]=2.35, while \citetalias{Marino2012} determined 1.4 dex. The thick solid line shows the synthetic spectrum that results using \citetalias{Marino2012}'s abundances adjusted using the offsets described in Section \ref{sec:Results} and Figure \ref{fig:CFe_M12_+03}. The thin solid line is the observed spectrum from \citetalias{vanLoon2007} and the dotted line is our fit to this spectrum.\label{fig:36179}}.
\end{figure}

A comparison of our [C/Fe] and [N/Fe] to that of \cite{Marino2012} showed that there was a systematic offset between the two datasets. In order to place this work onto the same abundance scale as \citetalias{Marino2012} it was necessary to apply some systematic corrections to our values. Using the eight of the nine stars in common (LEID 36179 was excluded as an outlier), it was found that the [C/Fe] needed the addition of $+0.45$ dex, while the offset in [N/Fe] was $-0.55$ dex. This offset was likely caused by continuum placement and the positive feedback between [C/Fe] and [N/Fe]. If the [C/Fe] is too low, then the [N/Fe] will have to be increased in order to reach the same strength of the CN region. The comparison of the corrected [C/Fe] and [N/Fe] values are shown in Figure \ref{fig:CFe_M12_+03}.

The significant outlier in Figure \ref{fig:CFe_M12_+03} is LEID36179. Figure \ref{fig:36179} shows the synthetic spectra produced using our parameters and the parameters of \citetalias{Marino2012} for this star. This clearly shows that the \citetalias{Marino2012} C and N abundances do not provide a good fit to the CN and CH spectral features in the \citetalias{vanLoon2007} spectrum. One note about this star is that \citetalias{Marino2012} determined an [O/Fe]$=-0.1$ while \citetalias{Johnson2010} have [O/Fe]$=-0.42$. This is the largest difference for these nine stars in terms of the oxygen abundance determined by the two groups. The overall results in this research are not compromised by this outlier.

\begin{figure}
\includegraphics[width=84mm]{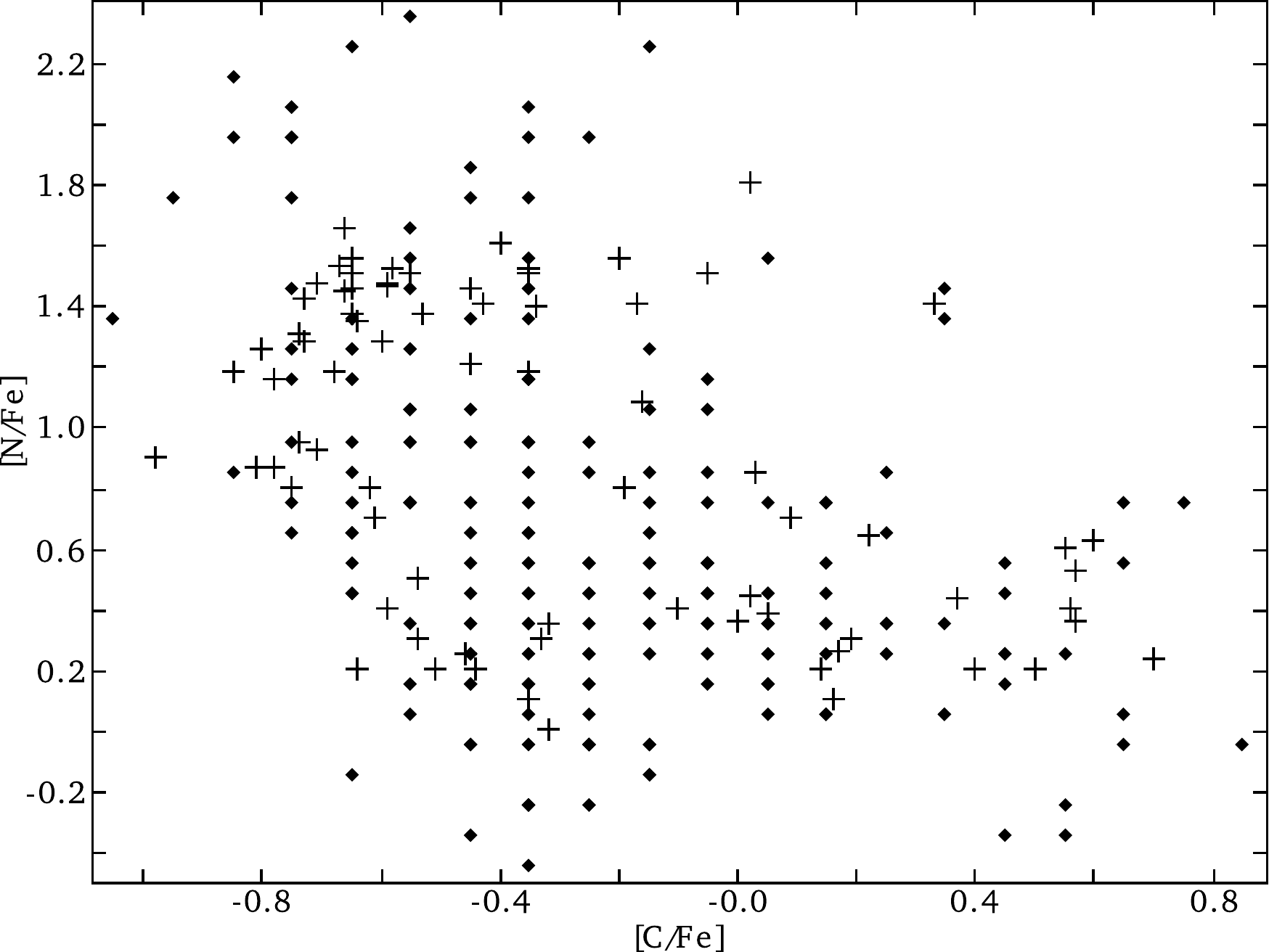}
\caption{The [C/Fe] and [N/Fe] for this work (filled diamond) and the \citetalias{Marino2012} $(+)$. The abundances were only determined to a precision of 0.1 dex due to the spectral resolution of the data. It shows the expected anticorrelation, as observed by \citetalias{Marino2012}. This plot excludes ten stars that had extremely low [N/Fe] abundances. \label{fig:CFe_NFe_M12_comp}}
\end{figure}

A qualitative comparison (Figure \ref{fig:CFe_NFe_M12_comp}) of the [C/Fe] to [N/Fe] shows that the automated spectral matching method worked well: it did not find stars that were both carbon and nitrogen rich, the upper-right quadrant on that plot. There was not the clear bifurcation that was seen by \citetalias{Marino2012}, but there was clearly the same anti-correlation between [C/Fe] and [N/Fe]. Of interest were the very nitrogen-rich, carbon-poor stars that are outside the range found by \citetalias{Marino2012}. These could be related to the way the spectral matching worked. It kept searching for the parameters that minimized the $\chi^2$, which could sometimes lead to extremely large or small values as the $\chi^2$ difference between the steps in [C/Fe] or [N/Fe] were very small. This was the cause of the extremely low [N/Fe] values that are not shown on Figure \ref{fig:CFe_NFe_M12_comp}. These stars have so little detectable nitrogen that that the automated process determined values as low as [N/Fe]$=-3$. This combination of the parameters and the resolution gives just an upper limit on the [N/Fe], rather than a definitive abundance.

\section{Clustering analysis}\label{sec:ClusteringAnalysis}

\begin{table*}
\begin{minipage}{155mm}
\caption{The cluster centres found using $k$-means clustering analysis of the 221 stars from the \citetalias{vanLoon2007} spectral library. For each parameter, the average and standard deviation are given, along with the number of stars in each group. The symbol is the identifier used for that group in the figures.}
\label{Table:Simpson2012Stats}
 \begin{tabular}{rcrrrrrrr}
 \hline
 Group & Symbol &  N. stars & [Fe/H] & [C/Fe] & [N/Fe] & [O/Fe] & [Na/Fe] & [Ba/Fe] \\
 \hline
1 & \textasteriskcentered & 62 & $-1.8\pm0.3$ & $-0.4\pm0.2$ & $0.3\pm0.3$ & $0.3\pm0.1$ & $-0.1\pm0.3$ & $-0.4\pm0.5$\\
2 & \textcolor{squares}{$\square$} & 86 & $-1.7\pm0.2$ & $0.0\pm0.3$ & $0.4\pm0.3$ & $0.4\pm0.1$ & $0.0\pm0.2$ & $0.4\pm0.3$\\ 
3 & \textcolor{triangles}{$\blacktriangle$} & 49 & $-1.7\pm0.3$ & $-0.5\pm0.2$ & $1.0\pm0.3$ & $-0.3\pm0.4$ & $0.3\pm0.2$ & $0.5\pm0.4$\\ 
4 & \textcolor{crosses}{$\times$} & 24 & $-1.3\pm0.4$ & $-0.6\pm0.2$ & $1.8\pm0.4$ & $-0.2\pm0.2$ & $0.7\pm0.2$ & $0.8\pm0.4$\\
\hline
 \end{tabular}
\end{minipage}
\end{table*}

\begin{table*}
\begin{minipage}{155mm}
\caption{The cluster centres for the 74 stars from the \citet{Marino2011,Marino2012} data. The column descriptions are the same as those in Table \ref{Table:Simpson2012Stats}.}
\label{Table:Marino2012Stats}
 \begin{tabular}{rcrrrrrrr}
 \hline
 Group & Symbol     & N. stars & [Fe/H] & [C/Fe] & [N/Fe] & [O/Fe] & [Na/Fe] & [Ba/Fe] \\
 \hline
        
1 & \textasteriskcentered & 17 & $-1.9\pm0.1$ & $-0.5\pm0.3$ & $0.5\pm0.3$ & $0.3\pm0.1$ & $0.1\pm0.2$ & $-0.4\pm0.2$\\
2 & \textcolor{squares}{$\square$} & 18 & $-1.6\pm0.2$ & $0.3\pm0.2$ & $0.4\pm0.2$ & $0.6\pm0.2$ & $0.1\pm0.1$ & $0.4\pm0.2$\\ 
3 & \textcolor{triangles}{$\blacktriangle$} & 19 & $-1.6\pm0.1$ & $-0.6\pm0.2$ & $1.3\pm0.2$ & $0.0\pm0.2$ & $0.4\pm0.2$ & $0.3\pm0.2$\\ 
3b & \textcolor{triangles}{$\triangle$} & 13 & $-1.5\pm0.1$ & $-0.5\pm0.2$ & $1.4\pm0.1$ & $-0.1\pm0.2$ & $0.6\pm0.1$ & $0.9\pm0.2$\\ 
4 & \textcolor{crosses}{$\times$} & 7 & $-1.1\pm0.1$ & $-0.3\pm0.3$ & $1.6\pm0.1$ & $0.2\pm0.2$ & $1.0\pm0.1$ & $0.8\pm0.1$\\ 
\hline
 \end{tabular}
\end{minipage}
\end{table*}

$k$-means clustering is a method to create $k$ groups from a dataset of $n$ entries using $m$ parameters. It minimizes the distance of each member from the group (or cluster) centre. In the work presented here, the method was implemented using the $R$ statistical package \citep{R2011}, a freely available system for statistical computation and graphics. The default $R$ algorithm of \citet{Hartigan1979} was employed in this work.

There are two caveats for $k$-means clustering. First, that the Euclidean distance is used for determining the clustering. This could require the observational data be transformed to give an appropriate range of values. This was not done in this work as the range of values was small and of similar magnitudes. Second, the number of groupings must be appropriate for the dataset. Too few or too many groupings can lead to less definitive results. In this case, there was experimentation with a range of the number of clusters.

\citet{Gratton2011} used $k$-means clustering on the \citetalias{Johnson2010} dataset to create seven groupings of stars in $\omega$~Cen. They used the [Fe/H], [$\alpha$/Fe], [Na/O] and [La/Fe] to show there are three metallicity regimes in the cluster with a range of $\alpha$, light element and neutron capture properties. Their analysis confirmed each metallicity regime has its own correlations and anticorrelations amongst the different elements. Of particular interest was the very metal-rich group that showed a Na-O \textbf{correlation} and had the highest average neutron-capture abundances of the groups in the cluster they identified.

We recognized that our low resolution data had larger uncertainties in each parameter than \citetalias{Johnson2010}. This would make group identification harder. In order to first understand the groups that could be found when carbon and nitrogen were used in conjunction with some of the abundances that \citet{Gratton2011} used, we decided to combine the \citet{Marino2011,Marino2012} datasets. There were 74 stars which had an [Fe/H] and [C,N,O,Na,Ba/Fe]. These six parameters were chosen to look for groupings that could be also found in our dataset of 221 stars. Neither \citet{Marino2011} nor \citetalias{Marino2012} determined an $\alpha$-element abundance. However, \citet{Gratton2011} found that there was a positive correlation between metallicity and their $\alpha$-element average for their groups.

In our analysis of \citetalias{Marino2012}, it was found that a clustering analysis of five groups gave the best results. Two groups would simply split between Na-strong and Na-weak stars. Four groups did not isolate the stars with extreme sodium abundances, with these stars being found in a group that consisted of a large range of metallicities and oxygen abundances but the same barium abundances. Although five groups were used, two of the groups were very similar in their properties so were combined into one group (\#3a \& \#3b).

The clustering analysis was then performed on our results using [Na,O/Fe] from \citetalias{Johnson2010}, and [C,N,Ba/Fe] and [Fe/H] from this work. Although we were able to define five groups from the \citet{Marino2011,Marino2012} data, only four groups were attempted with our data. This was because one of those groups (group \#3b) found in \citet{Marino2011,Marino2012} was primarily the result of it having higher barium abundance for its metallicity. This level of precision was not available in our data. As such, the same four overall groups were found with group \#3 having no subset that was s-process enhanced. Although \citet{Gratton2011} determined seven groups from the \citetalias{Johnson2010} dataset, we did not think that our smaller sample with lower precision abundances would allow us to define that many groups. Any further division created signal out of noise.

The cluster centres of each group are given in Tables \ref{Table:Simpson2012Stats} and \ref{Table:Marino2012Stats}. In the context of this paper, group \#1 will be referred to as metal-poor, groups \#2 \& 3 as intermediate metallicity and \#4 as metal-rich. By way of comparison, our group \#1 is the equivalent of groups 4 \& 6 in \citet{Gratton2011}, our \#2 is their 1 \& 5, our \#3 is their 2a \& 3, and our \#4 is their 2b.

\subsection{Groupings on the colour-magnitude diagram} \label{sec:CMD}
\subsubsection{\protect\citet{Marino2011,Marino2012}}\label{sec:MarinoAGB}
\begin{figure}
\includegraphics[width=84mm]{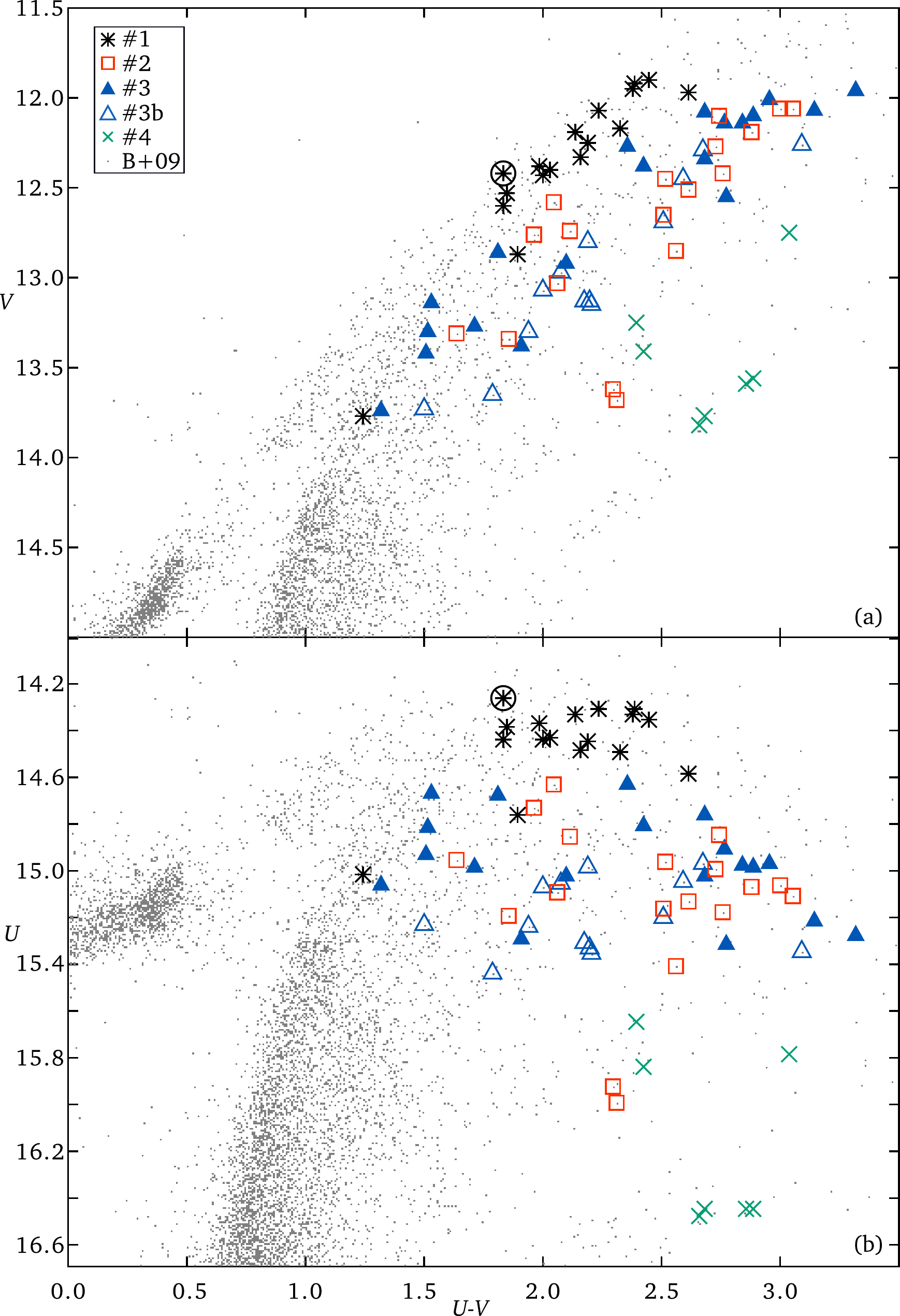}
\caption{(a) $V$-(\uv) and (b) $U$-(\uv) colour-magnitude diagrams of the \citetalias{Marino2012} stars, with different symbols for the different groups as described in Table \ref{Table:Marino2012Stats}. The small dots are the whole \citet{Bellini2009a} dataset. The four stars with the faintest $U$ magnitude are part of the metal-rich group and are on the RGB-a. The AGB star described in Section \ref{sec:MarinoAGB} is circled. \label{fig:CMD_Groups}}
\end{figure}
\begin{figure}
\includegraphics[width=84mm]{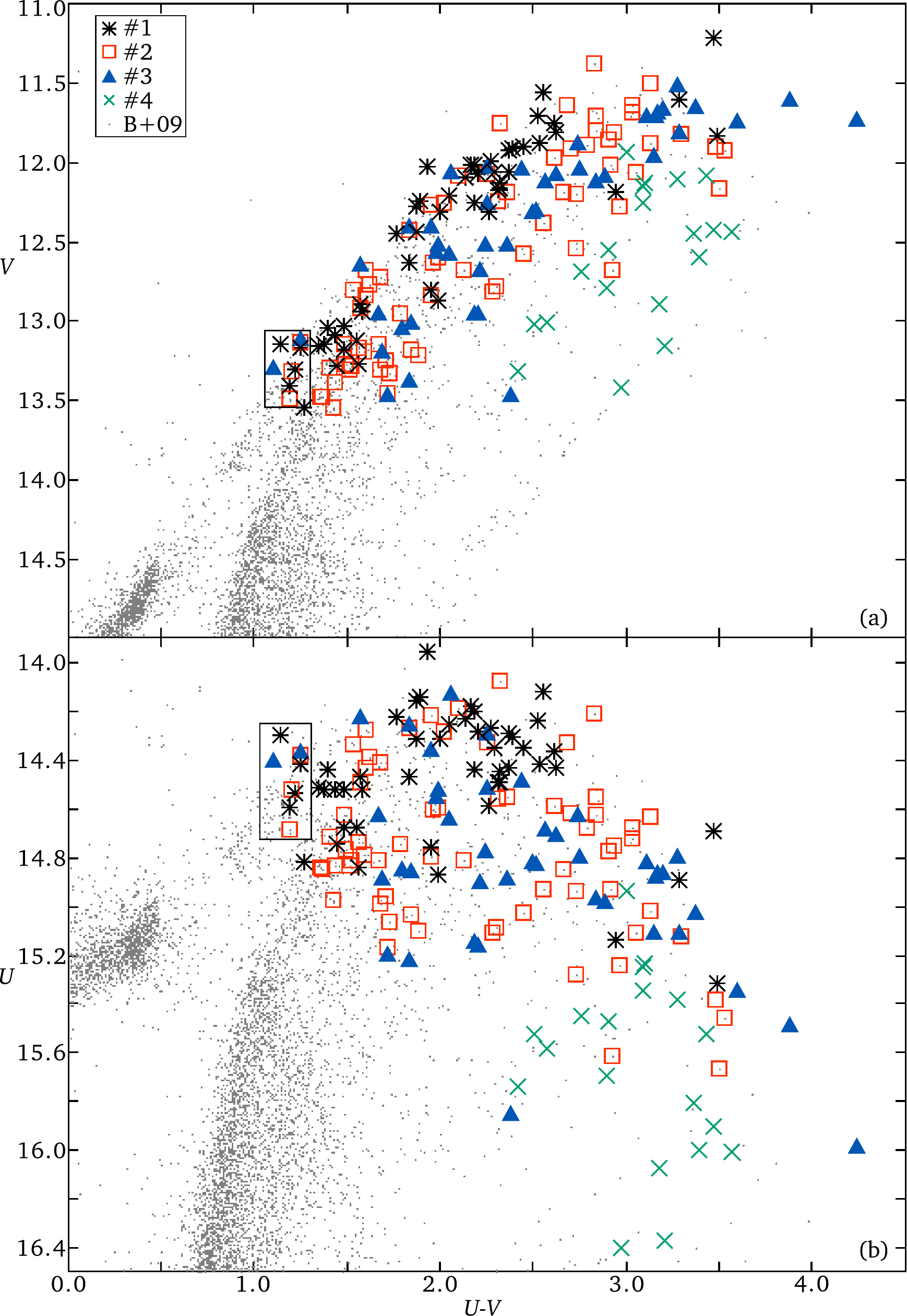}
\caption{The same as Figure \ref{fig:CMD_Groups} but using the 221 stars from this study. Symbols are the same as in Table \ref{Table:Simpson2012Stats}. The sequences are not quite as neatly defined as with the \citet{Marino2011,Marino2012} data but do show the same effects. We find that the metal-poor group (\#1) is concentrated on the hot side of the giant branch and the metal-rich stars (group \#4) are found in the anomalous RGB sequences. There are a number of potential and probable AGB stars found. The nine stars that are between $13.0>V>13.5$ and $(\uv)<1.4$ are highly likely to be AGB stars (surrounded by a rectangle). Only 172 of 221 stars are plotted on this diagram as not all the stars had $U$ photometry in the \citet{Bellini2009a} photometric library.  \label{fig:CMD_Groups_vL07}}
\end{figure}
The stars of \citet{Marino2011,Marino2012} were combined with \citet{Bellini2009a} and its high precision photometry. Figure \ref{fig:CMD_Groups} shows the $V$-(\uv) and $U$-(\uv) CMD with the five groups plotted. $U$ photometry is used here instead of $B$ as it was found to create a greater separation between the groups. Not all of the stars in our 221 star sample have $U$ photometry but there was no bias in the stars which did, and did not, have $U$ magnitudes determined by \citet{Bellini2009a}.

The metal-poor group is clearly on the hot-edge of the giant branch with little colour spread. As the metallicity increases, the stars are found at cooler positions. There are four stars in group \#4 which are likely members of the RGB-a \citep{Pancino2000}, the most metal-rich sequence in the cluster. These stars are found to be the most enriched in all the elements measured, except sodium where all of group \#4 are equally enhanced.

As would be expected from their metallicity ranges, groups \#2 and \#3 have very similar magnitude and colour ranges, while group \#3b is slightly cooler with a tight locus like group \#1. As discussed in the following sections, group \#3b could be an evolutionary extreme of group \#3 or the end point of the sodium-oxygen anticorrelation trend before the stars begin to be sodium-oxygen correlated.

The correlation between metallicity and colour is expected from stellar evolutionary models. The small spread in [Fe/H] in group \#1 correlates with a small spread on the CMD. Conversely, the larger spreads in the intermediate metallicity groups are shown on the CMD. On the \Teff-\logg\ plane, a straight line was fitted through the group \#1 stars, which could be used to produce a vertical plot of temperature deviation from the expected temperature for that gravity. The standard deviation of the temperatures in group \#1 using this technique was 30K. Group \#2 were $100\pm75K$ cooler, \#3 were $40\pm90K$, \#3b were $80\pm30K$ and \#4 were $300\pm60K$. 

It is possible that star ID 215367 (\citealt{Marino2011} numbering as this star is not in the \citetalias{vanLoon2007} library) is an AGB star (circled in Figure \ref{fig:CMD_Groups}), as it sits slightly to the left of the main locus of group \#1 stars on Figure \ref{fig:CMD_Groups}. This is most obvious in a ($U$,\uv) CMD (Figure \ref{fig:CMD_Groups}b), as this region of the GB is at an almost constant $U$ magnitude. Here this star is at a much brighter $U$ magnitude than would be expected for its \uv. This star is found to have the lowest [C/Fe] (and [C/H]) of all the stars measured (discussed in Section \ref{sec:CN} and shown in Figure \ref{fig:CFe_NFe_groups}). It is the most barium-rich of all the group \#1 stars (Section \ref{sec:sprocess}).

\subsubsection{This work}\label{Sec:Simpson2012CMD}
Plotting the colour-magnitude diagrams (Figure \ref{fig:CMD_Groups_vL07}) for our results has the same findings as with \citet{Marino2011,Marino2012}. Compared to Figure \ref{fig:CMD_Groups} there is more blending of the groups which is caused by the larger standard deviations in the average metallicities of the groups than for \citet{Marino2011,Marino2012}. However at the bright end of the giant branch, there is still a tight locus of the most metal-poor stars. The metal-rich stars are also the coolest as would be expected and the two intermediate metallicity groups are inseparable.

Our larger sample size has found nine stars that are in a well-defined AGB sequence. These stars come from the metal-poor and intermediate metallicity groups and are all found in the carbon- and nitrogen-poor quadrant of the CN diagram. These and other AGB stars will discussed in Section \ref{sec:AGB} in the context of the potential absence of CN-strong stars on the AGB. The \citetalias{Johnson2010} dataset did not extend to magnitudes fainter than $V=13.5$, which means there are only two members of the RGB-a metal-rich giant branch \citep{Pancino2000}.

\subsection{Sodium-oxygen anticorrelation}
\begin{figure}
\includegraphics[width=84mm]{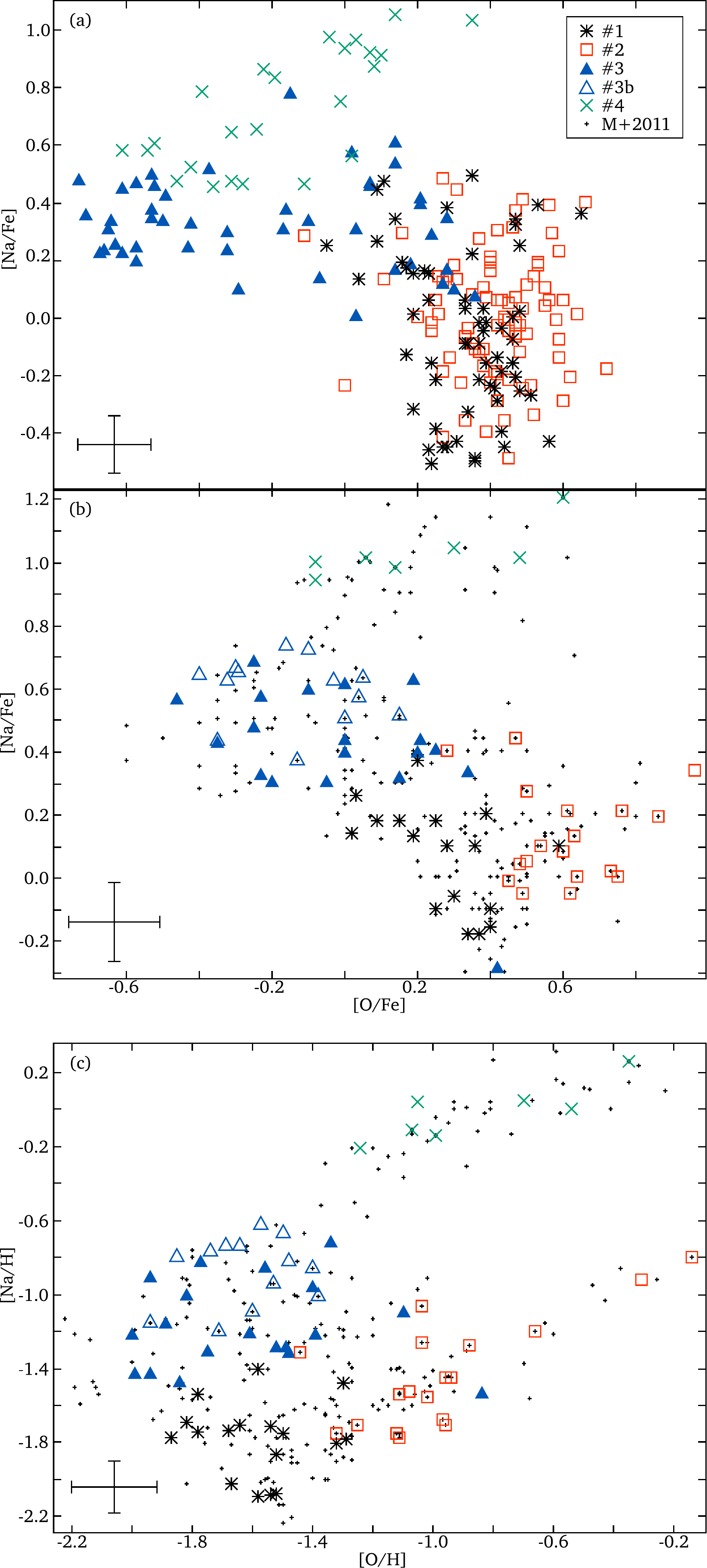}
\caption{The sodium-oxygen (anti)correlation for $\omega$~Cen of (a) this work and (b, c) \citet{Marino2011,Marino2012}. Symbols as described in Tables \ref{Table:Simpson2012Stats} \& \ref{Table:Marino2012Stats} and uncertainties are shown in the bottom-left of each diagram. The small dots in (b) \& (c) are the fuller \citet{Marino2011} dataset. They illustrate that features observed (for instance the correlations within group \#2 and \#4) are not due to small numbers of stars. The sodium-oxygen anticorrelation is clearly present between groups \#2 and \#3. There is little separation between the two subgroups of \#3, though potentially \#3b is slightly more sodium-rich than \#3. Group \#1 is found to be the most sodium-poor on average but with an intermediate oxygen, both with respect to iron and hydrogen. \label{fig:NaFe_OFe}}
\end{figure}
The $k$-means clustering of \citet{Marino2011,Marino2012} dataset split the Na-O anticorrelation at about [Na/Fe]$=0.3$, with groups \#1 \& \#2 being oxygen-rich, sodium-poor and groups \#3 \& \#4 having the opposite properties (Figure \ref{fig:NaFe_OFe}). The results broadly confirm the results found by \citetalias{Johnson2010} and \citetalias{Marino2012}. They both found that the intermediate metallicity stars displayed the well-known anticorrelation of mono-metallic clusters, while the metal-poor and -rich groups do not share this bimodal behaviour. The same result is observed for the clustering analysis of our data.

Plotting the abundances with respect to hydrogen instead of iron (Figure \ref{fig:NaFe_OFe}c), the group \#1 stars have the same [O/H] as group \#3 with [Na/H] different by $\sim1$ dex. Group \#2 stars are about 0.5 dex enhanced in [O/H] compared to group \#1 stars and potentially slightly enhanced in sodium. If group \#2 formed from the the pollution of group \#1 stars, this gas and dust would be enhanced in overall metallicity, oxygen and sodium. However if group \#3 were formed after group \#1, this would require an increase in metallicity and sodium and a depletion in oxygen.

\citet{D'Antona2011} would have groups \#3 and \#3b at opposite ends of the evolutionary sequence that forms the sodium-oxygen anticorrelation. They theorized that the first stars of the second generation were sodium-rich, oxygen-poor and formed from pure AGB ejecta. As this ejected material was diluted by mixing with the cluster's surrounding interstellar medium, progressively oxygen-richer, sodium-poorer stars would form. This would continue until the AGB ejecta from these stars becomes the majority contributor to the gas of the cluster and produces sodium-rich stars. In this way, \#3b could represent an extreme limit to sodium enhancement. The extension of group \#2 into extremely high oxygen values is seen in the \citet{Marino2011} dataset, but is not evident to the same extent in the larger \citetalias{Johnson2010} dataset. 

Both \citet{Marino2011} and our study show the \textbf{correlation} of sodium and oxygen in the most sodium-rich stars. This feature is not seen in mono-metallic clusters. The metal-rich stars of group \#4 form this sequence, which would imply that they are the youngest stars of this sample, forming from the ejecta of the last supernovae in the cluster. \citet{Gratton2011} found this group to be the most enhanced in $\alpha$-elements as well. The correlation is most dramatically seen in Figure \ref{fig:NaFe_OFe}a which used the \citetalias{Johnson2010} abundances. Here the correlation extends from [O/Fe]$=-0.6$ to $+0.6$.

\subsection{Carbon and nitrogen} \label{sec:CN}
\begin{figure}
\includegraphics[width=84mm]{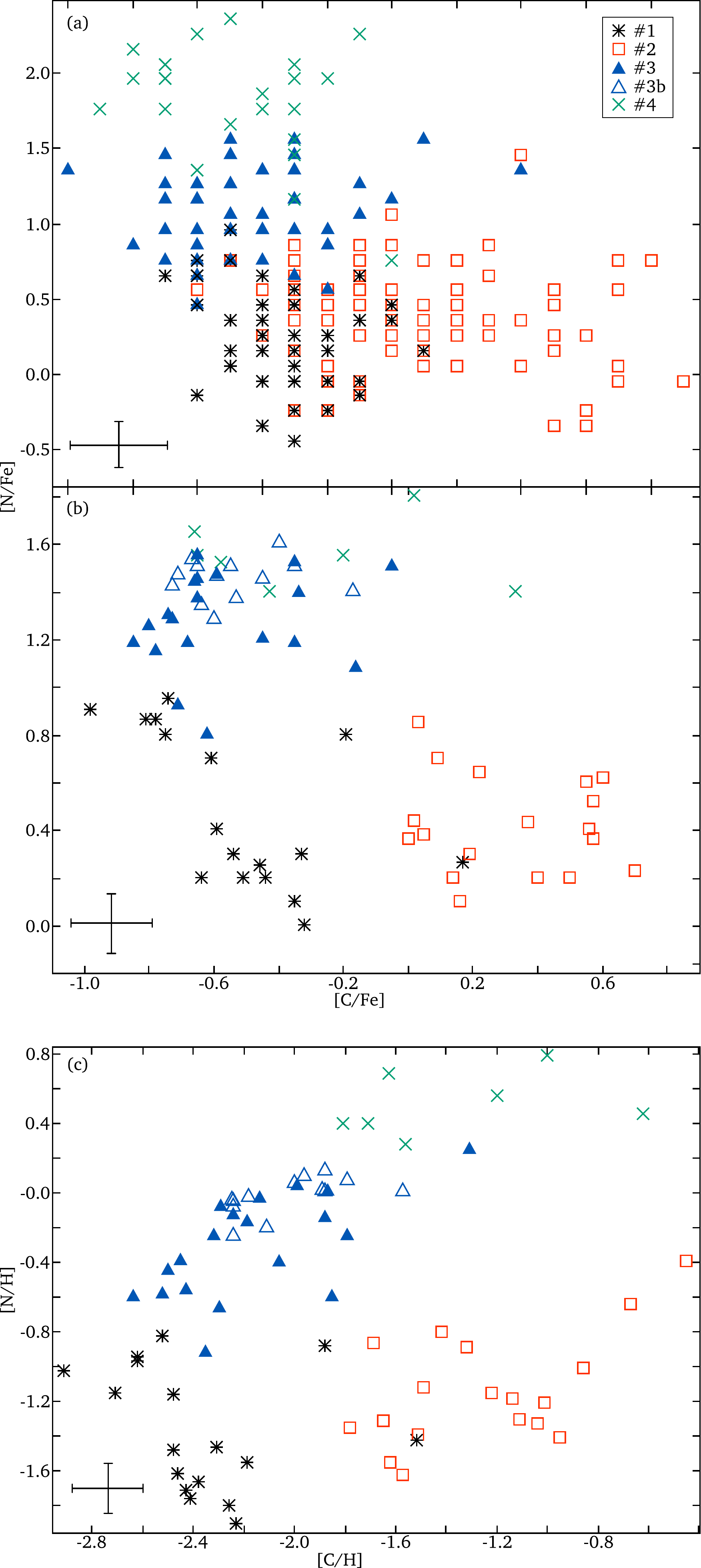}
\caption{Nitrogen against carbon for (a) this work and (b, c) \citet{Marino2011,Marino2012}. Uncertainties are shown in the bottom-left of each diagram. The C-N plane looks broadly similar to the Na-O plane. The metal-poor group (\#1) is found to be deficient in both carbon and nitrogen, while the intermediate metallicity groups are bimodal. Group \#3b shows an extreme enhancement of nitrogen.  When considered with respect to iron (a, b), group \#4 stars are very similar to those of group \#3 in terms of their carbon abundance. However when they are considered with respect to hydrogen (c), their high metallicities result in a separate sequence, with some up to 0.8 dex higher than the solar value in nitrogen. The potential AGB star (Section \ref{sec:MarinoAGB}) is the most carbon-deficient star in the \citet{Marino2011,Marino2012} sample (b, c). \label{fig:CFe_NFe_groups}}
\end{figure}
\begin{figure}
\includegraphics[width=84mm]{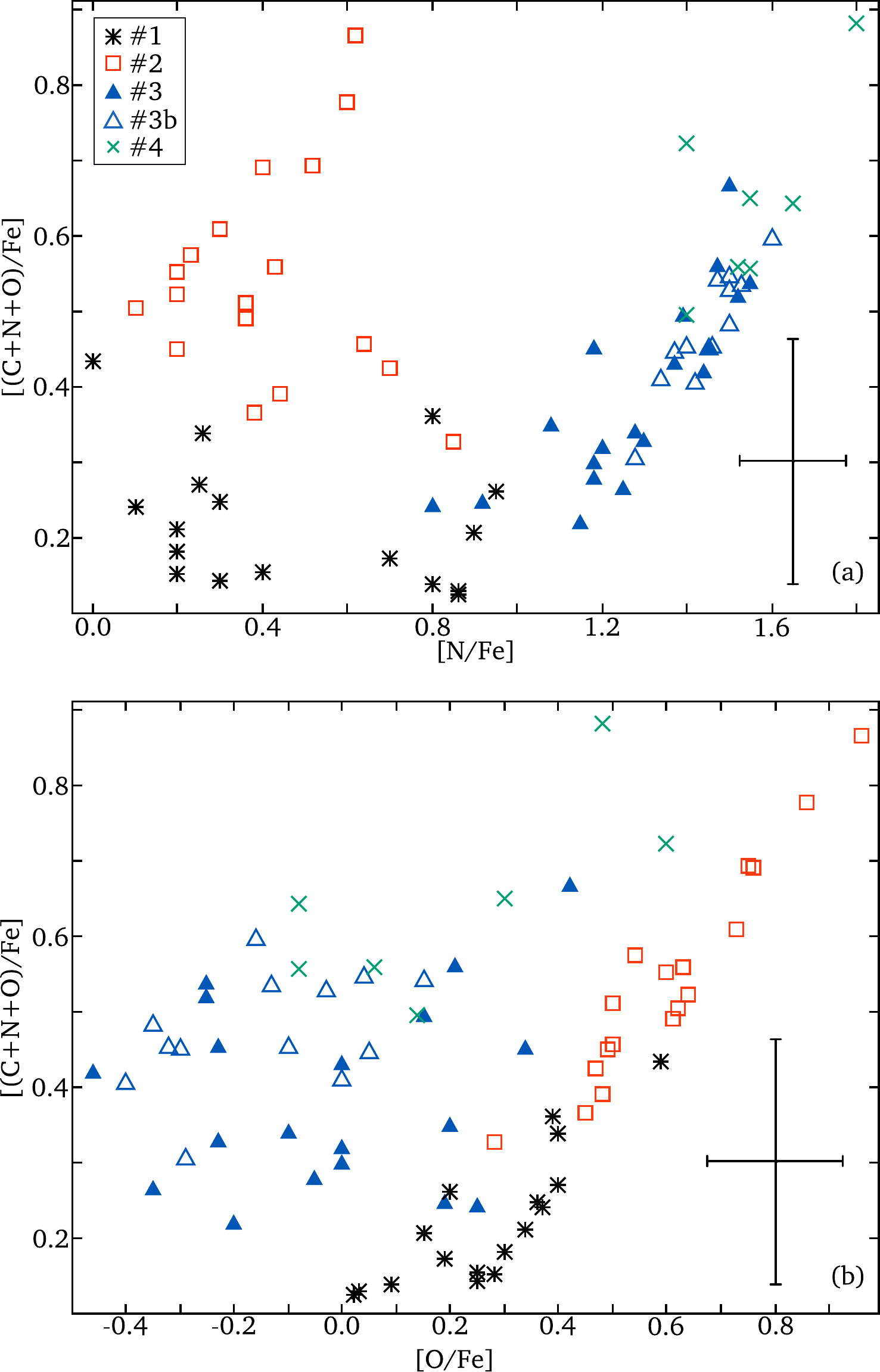}
\caption{The total C+N+O with respect to iron for the \citet{Marino2011,Marino2012} dataset \citep[using solar abundances from][]{Anders1989}. Symbols as described in Table \ref{Table:Marino2012Stats}. Uncertainties are shown in the bottom-right of each diagram. (a) the oxygen-rich groups (\#1 \& \#2) have oxygen as their dominant CNO species, (b) while in the nitrogen-rich groups (\#3 \& \#4) it is nitrogen. Unlike mono-metallic clusters, the total C+N+O is not constant. \label{fig:CNOFe_NFe}}
\end{figure}
As with sodium-oxygen, there are roughly three regions of chemical abundance for carbon and nitrogen (Figure \ref{fig:CFe_NFe_groups}): \#1 are C- and N-weak; \#2 are C-strong, N-weak; and \#3 and \#4 are C-weak, N-strong. There is a range of abundances in each grouping with \#1 showing a range of 1 dex in [N/Fe] and about 0.6 dex in [C/Fe]. The intermediate metallicity groups have a range of 0.6 dex in [N/Fe] and [C/Fe]. Group \#4, which has very few stars, ranges from [C/Fe]$=-0.65$ to $+0.3$. 

Groups \#3 and \#3b have similar ranges of [C/Fe] (Figure \ref{fig:CFe_NFe_groups}b) but the \#3b group form an upper limit for nitrogen. This is reinforced when the stars are plotted with [C/H] versus [N/H] (Figure \ref{fig:CFe_NFe_groups}c). The group \#3b stars have an average [N/H]$=0.01\pm0.06$ suggesting a maximum possible pollution for stars of this generation. Group \#4 stars are found to have even higher nitrogen abundances with the four RGB-a \citep{Pancino2000} stars having [N/H]$>0.45$ ([N/Fe]$>1.4$).

In mono-metallic clusters, the total C+N+O has been found to be constant \citep[e.g.][]{Carretta2005}. In the case of $\omega$~Cen there is a large spread (Figure \ref{fig:CNOFe_NFe}). It was found that groups \#1 and \#2 have a straight-line relationship between [O/Fe] and [C+N+O/Fe] but with the other two groups having no correlation. Between [N/Fe] and [C+N+O/Fe], it is instead groups \#3 and \#4 that have the positive correlation. This illustrates that in the first two groups, it is nitrogen which is the dominant CNO species, while in the other two groups it is oxygen. This is what would be expected based upon their relative abundances of nitrogen and oxygen.

In \citet{Marino2011,Marino2012} (see our Table \ref{Table:Marino2012Stats}) there are 39 (53\%) stars in the N-rich groups and 35 (48\%) in the N-poor group, while in our work (see our Table \ref{Table:Simpson2012Stats}) there are 148 (67\%) N-rich stars and 73 (33\%) N-poor stars. We attribute this to a selection effect. The split in nitrogen groupings maps to the split in oxygen. In the \citet{Gratton2011} analysis, they had 574 (72\%) stars in O-poor groups (N-rich) and 223 (28\%) stars in O-rich groups (N-poor). So our groupings follow that of \citet{Gratton2011} in terms of the split of N-rich and N-poor stars. Our selection of stars was effectively based upon the selection of \citetalias{Johnson2010} through the use of their [Ca/Fe] and [O/Fe] abundances.

\subsection{s-process}\label{sec:sprocess}
\begin{figure}
\includegraphics[width=84mm]{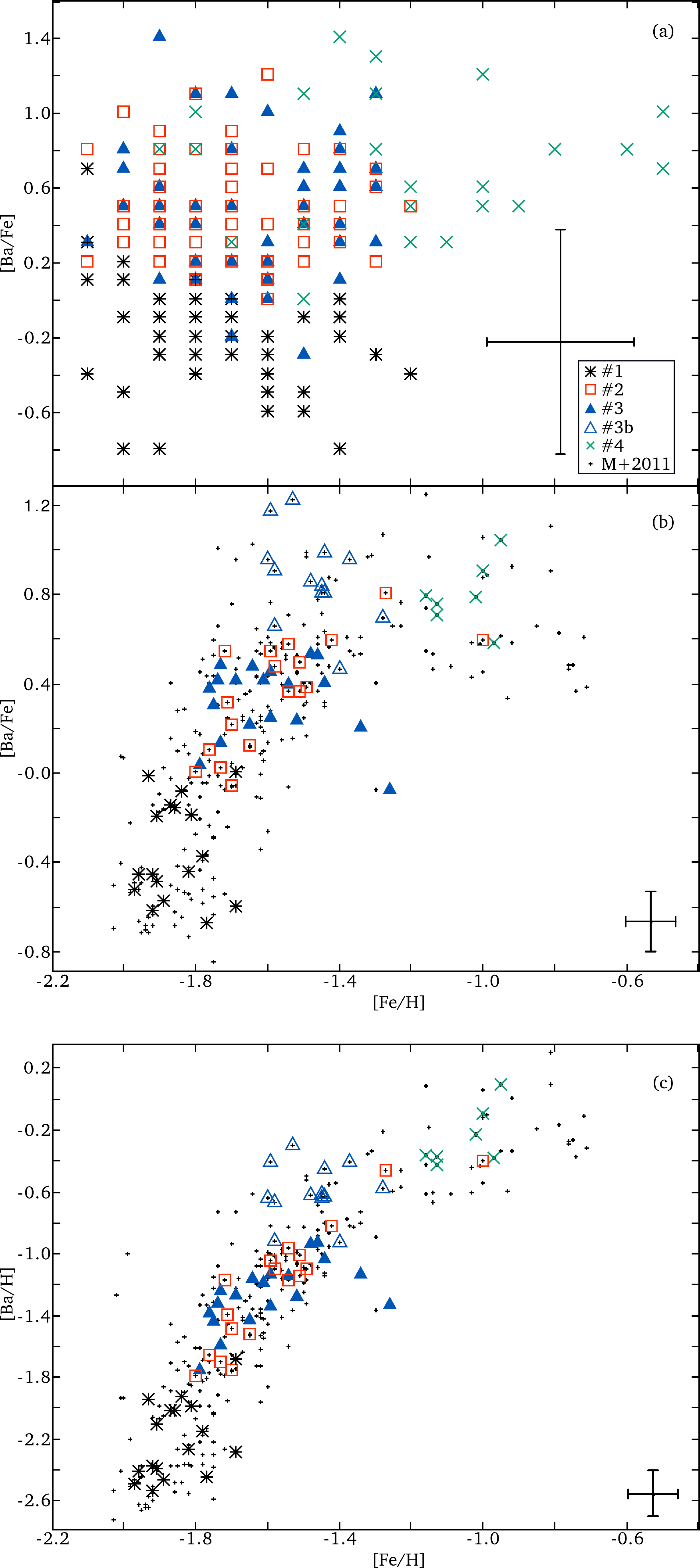}
\caption{The correlation between the metallicity of the stars and their s-process abundance for (a) this work and (b, c) \citet{Marino2011,Marino2012}. Uncertainties are shown in the bottom-right of each diagram. Only group \#4 does not form part of the correlation, with the s-process abundance constant with respect to iron. One of the causes of the separation of group \#3b from group \#3 is the roughly 0.5 dex difference in their [Ba/Fe]. \label{fig:BaFe_FeH}}
\end{figure}

A large range of s-process abundances is not typically observed in mono-metallic clusters. However, $\omega$~Cen does exhibit a very large range of abundances in elements such as barium. With a simple correlation between [Fe/H] and [Ba/Fe], the groupings were as would be expected from previous work. The metal-poor stars all had sub-solar abundances of barium with respect to iron, but with a very large range. Some stars had [Ba/Fe]$=-0.8$ while others had greater than the solar abundance in [Ba/H].

Unlike the other element combinations investigated in this study (C-N \& Na-O), barium does not exhibit any differentiation between the intermediate metallicity groups of \#2 and \#3. This could require that any enhancements in metallicity and s-process happened before the anticorrelations.

Here the separation of group \#3 from \#3b was most evident (Figure \ref{fig:BaFe_FeH}b). The two groups had about the same metallicity range but a difference of about 0.5 dex in [Ba/Fe]. A caveat to this is that in \citet{Marino2011} there were few stars at this metallicity/s-process abundance combination. They comment that at high barium abundances the lines get too strong for accurate measurement. However we have kept them as a separate grouping due to their separation in the other abundance planes that we investigated.

Neither \citet{Marino2011,Marino2012} nor this work determined any r-process abundances. \citetalias{Johnson2010} did determine [Eu/Fe]. As noted by \citet{Gratton2011} there is no trend of [Eu/Fe] with [La/Fe] in \citetalias{Johnson2010}, indicating that the contribution of the r-process to neutron-capture elemental abundances in the cluster are negligible. It would appear that there is a potential increase in [Eu/H] at high [La/H] and that the [La/Eu] abundance is dominated by changes in La. There is a larger scatter in [La/Fe] in the the metal-intermediate and -rich groups (almost 1 dex) but overall, in these groups, this ratio is constant.

\section{Lack of CN-strong AGB stars}\label{sec:AGB}
\begin{figure}
\includegraphics[width=84mm]{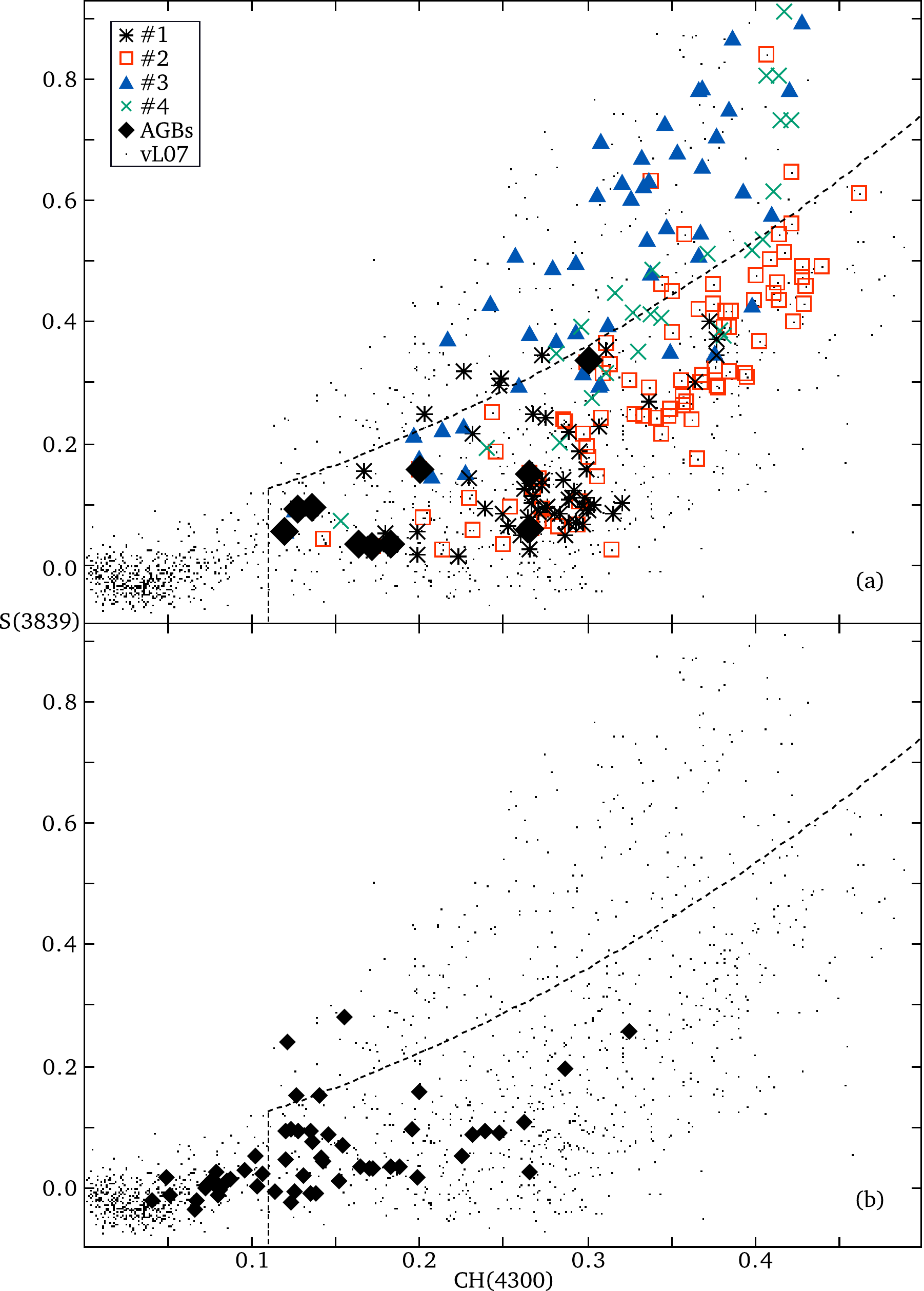}
\caption{(a) The four groups with their CN [S(3839)] and CH [CH(4300)] indices as found by \citetalias{vanLoon2007}. The large diamonds are the AGB stars identified in Section \ref{Sec:Simpson2012CMD}. The curved line separates the oxygen-rich groups, defined as [O/Fe]$>0.17$ (below the line) from the oxygen-rich stars. This serves as a divider between CN-weak and CN-strong stars. The vertical dashed line divides the HB from the GB in these diagrams. The small dots in all subplots show all the stars in the \citetalias{vanLoon2007} dataset. (b) Highlighting all the 55 AGB stars identified in the entire \citetalias{vanLoon2007} library based upon their position in the CMD of \citet{Bellini2009a}. \label{fig:vL07_Fig13}}
\end{figure}

\citet{Campbell2010} found evidence that in some globular clusters, there are no CN-strong stars on the AGB. This was first found by \citet{Norris1981} in NGC~6752. This lack of CN-strong stars on the AGB is in contrast to the RGB where the stars are equally CN-weak and CN-strong in all clusters, including $\omega$~Cen. In the case of \citetalias{Marino2012} there is only one potential AGB star. This star has a low [N/Fe] and [C/Fe], which would fit with the findings of \citet{Campbell2010}. From Section \ref{Sec:Simpson2012CMD}, there are nine potential AGB stars in the subset of \citetalias{vanLoon2007} for which [C/Fe] and [N/Fe] were determined.

\citetalias{vanLoon2007} measured two indices of interest: the S(3839) and CH(4300) indices\footnote{It should be noted that these values used \citetalias{vanLoon2007} normalized spectra.}. The first measured the CN band around 3839\AA\ and the second CH band at 4300\AA\ (the G band). Using the definition of \citet{Harbeck2003}, where $F_{\lambda1-\lambda2}$ is the average flux level between the two wavelengths in angstroms:
\begin{equation}
\rm{S(3839)}=-2.5\log\left(\frac{F_{3861-3884}}{F_{3894-3910}}\right),
\end{equation}
and
\begin{equation}
\rm{CH(4300)}=-2.5\log\left(\frac{2F_{4285-4315}}{F_{4240-4280}+F_{4390-4460}}\right).
\end{equation}

Figure \ref{fig:vL07_Fig13}a shows the S(3839) and CH(4300) indices strength for all the stars determined, with a curved line dividing the O-rich stars ([O/Fe]$>0.17$) (Figure \ref{fig:NaFe_OFe}) from the O-poor stars. This line takes the form of a parabola. The clump of stars in the lower-left quadrant of both subfigures of Figure \ref{fig:vL07_Fig13} with CH(4300)$<0.9$ are HB stars. This has been separated from the GB by a vertical dashed line. The groups of stars which were deficient in nitrogen (see Figure \ref{fig:CFe_NFe_groups}) are almost all below the dividing line, while group \#3 (C-weak, N-strong) stars are placed above the line. The (metal-rich) group \#4 are found to be on both sides of the line. For this group there is a much larger variation in S(3839) than in CH(4300), which would correspond to their larger range of [N/Fe] than [C/Fe]. The nine AGB stars have small to medium strength CN and CH bands based upon their strengths of their indices.

All this gives good evidence that we can use the S(3839) and CH(4300) indices of \citetalias{vanLoon2007} to define the CN strength of the stars. To extend our sample of AGB stars from just those for which a [C/Fe] and [N/Fe] was determined, the photometry of \citet{Bellini2009a} was used to separate the AGB from the RGB from $12.6<V<14.2$. By eye, over 200 AGB stars were defined from the \citet{Bellini2009a} sample (out of over 60,000 probable cluster members). These stars then gave 55 AGB stars in the overall \citetalias{vanLoon2007} sample for which there was also \citet{Bellini2009a} photometry.

All the \citetalias{vanLoon2007} stars identified as AGB stars have low S(3839) and CH(4300) indices (Figure \ref{fig:vL07_Fig13}b). In the case of the CN index, they are all less than 30\% of the maximum value measured for any star by \citetalias{vanLoon2007}. The vertical line places 18 of the 55 stars (33\%) in the HB part of the diagram. Some of these stars may not be HB stars but are too hot to easily determine their CN strength from these indices. Of the remaining 37, only two of these AGB stars (5\%) are significantly above the line. From this we would infer that more than 90\% of this AGB are CN-weak.

It should be noted that in the case of $\omega$~Cen, this refers to one particular AGB. Due to the range of ages, metallicities and helium content, there will be potentially as many distinct AGBs as there are distinct RGBs. This means that it is almost certain that some of the stars that are being referred to here, and in other works as RGB stars, will in fact be AGB stars. This level of contamination will also be affected by how many of stars actually ascend the AGB. \citet{DCruz2000} estimated that as many as 30\% of the horizontal branch stars are extreme in their colour, which could mean that they evolve directly to white dwarfs, never passing along the AGB.

In the case of the AGB identified in this work, the assumption would be that it is the AGB is associated with the hottest and most metal-poor RGB. This RGB is thought to be not helium-rich, being associated with the red main sequence. A low helium abundance would not support the theory of \citet{Norris1981} that the lack of CN-strong stars was the result of them being helium-rich and therefore evolving directly from the HB to the white dwarf phase.

\section{Discussion}\label{sec:EvolutionaryModels}
The clustering analysis identified four groups of stars, which would have had multiple star formation events within them. Each grouping showed a range of abundances in every element that was incorporated into the $k$-means clustering.

Our primordial population is group \#1. These stars are deficient in every element except oxygen. This implies that the material from which the cluster formed had not been polluted with material from AGB stars, if it is assumed that all generations of the cluster formed in the same spatial location and not as the result of a merger. Within this first generation there is still a metallicity range, greater by an order of magnitude than what is observed in mono-metallic globular clusters.

\citet{D'Antona2011} used the yields of massive AGB stars to model the sodium-oxygen anticorrelations (and correlation in the case of the very metal-rich stars) in $\omega$~Cen. Using the 300 stars of \citet{Marino2011} they identified that for all the stars in the metallicity range of $-1.9<{\rm [Fe/H]}<-1.3$, there was a Na-O anticorrelation, as observed in our groups \#2 \& \#3. This implies that the process through which the anticorrelations formed will be broadly similar to that which formed the anticorrelations in mono-metallic clusters.

The most metal-rich stars in the cluster ([Fe/H]$>-1.3$) show a direct sodium-oxygen correlation (Figure \ref{fig:NaFe_OFe}). This, \citet{D'Antona2011} suggested, would be the result of stars forming from the pure ejecta of super-AGBs. In these stars the sodium and oxygen yields are correlated with the yields of both elements decreasing with stellar mass to about 6.5~$M_{\sun}$, then the yields begin to increase again. This is caused by the changing temperature of the convective envelope in these stars. By the time the most metal-rich stars in the cluster form, this AGB material would not be diluted as any residual ISM would have been blown away and/or consumed in preceding star formation episodes. This requires that future star-forming events would form from just the ejecta of AGB stars. Our analysis finds that these most metal-rich stars, with their high oxygen and sodium, are nitrogen-rich, but with no large enhancement of the s-process abundances found for their metal-poorer predecessors at high sodium but low oxygen.

In figure 3 of \citet{D'Antona2011}, there is a possible scenario for the production of sodium-oxygen (anti)correlation(s) in a mono-metallic cluster (in their case NGC 2808). Using the model of \citet{DErcole2008,DErcole2010} they used super-AGB star yields to show a qualitative match to the Na-O anticorrelation and correlation of $\omega$~Cen. In this model, the first stars of the second generation are the most oxygen-poor, forming from pure AGB ejecta which has accumulated at the centre of the cluster. These stars will be very helium-rich (corresponding to the blue main sequence). This pure ejecta is then diluted by ``pristine'' gas from the surrounding interstellar medium, which reduces its sodium abundance and increases its oxygen abundance.

Of course such a model cannot be directly applied to $\omega$~Cen with its large metallicity range. The ``pristine'' gas that dilutes the pure AGB gas must be already enhanced in metallicity and s-process elements. The carbon-nitrogen anticorrelation also requires that this ``pure'' AGB gas be enhanced in nitrogen with no corresponding enhancement of carbon. Then the dilution with pristine gas will result in depletion in nitrogen and the enhancement in carbon.

One of the most curious aspects of $\omega$~Cen is the correlation between metallicity and s-process abundance. Here we found that both aspects of the Na-O and C-N anticorrelation in the metal-intermediate groups showed the same range of s-process abundance. This would imply that any s-process enhancement that these intermediate metallicity, second generation stars experienced must have taken place before  the second generation of star formation took place.

This s-process correlation has always been difficult to reconcile with other aspects of the chemical evolution of the cluster. The generally accepted location of the formation of s-process elements is low-mass AGB stars (M$<3M_{\sun}$). These stars require a timescale of up to a gigayear to evolve, while the source of the sodium-oxygen anticorrelation, higher-mass AGBs (M$>5M_{\sun}$), will have already evolved. As such there must be a substantial delay in the star formation that took place to create the second generation of stars. And then a subsequent delay to form the most metal-rich stars, which again are even more enhanced in s-process. The most metal-rich stars are however not enhanced to the same extent, with [Ba/Fe] at the same values as the most-enhanced stars of the second generation.

A clue to the formation of this cluster is that the oxygen-poor stars are centrally concentrated. \citetalias{Johnson2010} found that 91\% of the stars with [O/Fe]$\leq0$ were within 10 arcminutes of the cluster centre in their sample. This would be our groups \#3 and \#4. They inferred that the oxygen-poor stars were also from the blue main sequence (bMS), which is also the helium-rich sequence. So this implies that the oxygen-rich, helium-rich stars are also nitrogen-rich, carbon poor. 

This matches with the theory of \citet{DErcole2008,DErcole2010} that the oxygen-poor, helium-poor stars were the first of the second generation stars to form and did so with a central concentration as this was where the cooling flows brought the pure AGB gas. Their theory would then require that the oxygen-rich stars that form later will have no radial gradient, instead forming throughout the cluster. 

An alternative theory was proposed in the recent paper of \citet{Herwig2012}. They identified galactic plane passages as mechanism for stripping the cluster of gas. Their evolutionary sequence would have our group \#1 as the primordial population, followed by \#3 forming and then \#4. The last population formed would be \#2 from the gas of \#1 after all any residual gas had been stripped by a galactic plane passage.

One key element that requires measurement for a large number of stars in $\omega$~Cen is helium. Do our group \#2 differ greatly in their helium abundance from group \#3 and do the group \#3 subsets have differences in helium as well? The most recent determination of helium abundances was by \citet{King2012} who used isochrones to determine that the bMS had a $Y = 0.39 \pm 0.02$. \citet{Dupree2011} directly measured helium in giant branch stars and found that those stars with high sodium and aluminium tended to show helium, while those with [Na/Fe]$<0.2$ did not show helium in their spectra. In terms of the data presented here, this would be groups \#3 and \#4 (and the most sodium-rich stars of groups \#1 and \#2) which would potentially show helium in their spectra. So the polluters that formed that gas and dust for group \#3 would also be enhancing the helium abundance of these stars. This fits into the model hypothesized by \citet{DErcole2008,DErcole2010}. 

\citet{Sbordone2011} calculated synthetic spectra based upon typical globular cluster abundances. These spectra were then used to create synthetic CMD for a variety of bandpasses, one of which was the $U$ and $V$ used in this work. They noted that CNO variations will cause the greatest variation in the $U$ and $B$ bands due to the molecular lines in this region of the spectrum. Their base isochrone had a metallicity of $-1.62$ ($Z=0.001$, $Y=0.246$) with an age of 12 Gyr. Three other abundance combinations were investigated: one enhanced in nitrogen and sodium by 1.8 and 0.8 dex respectively and depleted in carbon and oxygen by 0.6 and 0.8 dex (CNONa1); one with the same enhancement in C, Na and O but only enhanced by 1.4 dex in nitrogen (CNONa2); and the same as the first but with $Y=+0.4$ (CNONa1+He). The C+N+O mass fraction was twice the reference value in CNONa1, while CNONa2 had the same value as the reference. This is similar to the hypothesized situation in $\omega$~Cen: our group \#3 has a larger C+N+O total than group \#2 (Figure \ref{fig:CNOFe_NFe}) and from \citet{DErcole2008,DErcole2010} group \#3 would be enhanced in helium. They found that in the $U-(\ub)$ CMD, the CNONa1+He and CNONa2 were indistinguishable for much of the RGB. This would concur with what we have found in this work: that groups \#2 and \#3 are blended on the CMD. But of course, their work used a mono-metallic isochrone and could not be fitted to the GB of $\omega$~Cen.

\section{Conclusions}
The \citetalias{vanLoon2007} spectral library provides a large sample of spectra of giant branch and horizontal branch stars in the globular cluster $\omega$~Cen. Combining this spectral data with high precision photometric data and previously determined abundances of these stars, we determined the [C/Fe] and [N/Fe] for over 200 new stars in the cluster. These results confirmed the results of \citetalias{Marino2012}: that the low metallicity stars have both low carbon and nitrogen, while the intermediate groups show an anticorrelation.

Combining this carbon and nitrogen data with sodium, oxygen and barium abundances, and using $k$-means clustering, we found four (potentially five) groupings of stars (with metallicity averages from Table \ref{Table:Marino2012Stats}):
\begin{itemize}
\item A low-metallicity group ($\langle$[Fe/H]$\rangle=-1.9$) which is deficient in carbon and nitrogen relative to iron, but not oxygen.
\item Two intermediate metallicity groups ($\langle$[Fe/H]$\rangle=-1.6$) which are anti-correlated in Na-O and C-N. Group \#2 are high in C and O, while group \#3 are high in Na and N. Both these groups have the same metallicity and barium range. There is a potential subgroup of group \#3, which was labelled \#3b which has a higher barium than the rest of the group and higher Na and N without any differences in its O and C.
\item A metal-rich group ($\langle$[Fe/H]$\rangle=-1.1$) which is enhanced in Na, N, and showing a correlation of Na with O. This group has the highest average barium of all the groups.
\end{itemize}

High precision photometry was used to define 39 AGB stars in the \citetalias{vanLoon2007} sample. Their S(3839) and CH(4300) indices provided a good measure of the CN strength of the star. The majority of AGB stars were found to be CN-weak. The hypothesis that differing helium abundances could cause such an effect in mono-metallic clusters does not fit this result, as this AGB is associated with the most metal-poor stars in the cluster and therefore will not have any helium enhancement.

\section*{Acknowledgments}
The authors were supported by the Marsden Fund Council from New Zealand Government funding, administered by the Royal Society of New Zealand. JDS was also funded by the University of Canterbury. The authors would like to thank the anonymous referee for their insightful and helpful comments that improved the manuscript. We also thank Andrew Ridden-Harper for the CN and CH matching code; Wolfgang Kerzendorf for a Python wrapper for MOOG; Erik Brogt for proof-reading and helpful article writing advice.

This publication makes use of data products from the Two Micron All Sky Survey, which is a joint project of the University of Massachusetts and the Infrared Processing and Analysis Center/California Institute of Technology, funded by the National Aeronautics and Space Administration and the National Science Foundation.

\bibliographystyle{mn2e}
\bibliography{references}

\begin{thebibliography}{}

\bibitem[\protect\citeauthoryear{{Allen} \& {Cox}}{{Allen} \&
  {Cox}}{2000}]{Allen2000}
{Allen} C.~W.,  {Cox} A.~N.,  2000, Allen's Astrophysical Quantities, 4th edn.
Springer, New York

\bibitem[\protect\citeauthoryear{{Alonso}, {Arribas} \&
  {Mart{\'{\i}}nez-Roger}}{{Alonso} et~al.}{1999}]{Alonso1999}
{Alonso} A.,  {Arribas} S.,    {Mart{\'{\i}}nez-Roger} C.,  1999, A\&AS, 140,
  261

\bibitem[\protect\citeauthoryear{{Anders} \& {Grevesse}}{{Anders} \&
  {Grevesse}}{1989}]{Anders1989}
{Anders} E.,  {Grevesse} N.,  1989, GeCoA, 53, 197

\bibitem[\protect\citeauthoryear{{Bedin}, {Piotto}, {Anderson}, {Cassisi},
  {King}, {Momany} \& {Carraro}}{{Bedin} et~al.}{2004}]{Bedin2004}
{Bedin} L.~R.,  {Piotto} G.,  {Anderson} J.,  {Cassisi} S.,  {King} I.~R.,
  {Momany} Y.,    {Carraro} G.,  2004, ApJL, 605, L125

\bibitem[\protect\citeauthoryear{{Bellini}, {Piotto}, {Bedin}, {Anderson},
  {Platais}, {Momany}, {Moretti}, {Milone} \& {Ortolani}}{{Bellini}
  et~al.}{2009}]{Bellini2009a}
{Bellini} A.,  {Piotto} G.,  {Bedin} L.~R.,  {Anderson} J.,  {Platais} I.,
  {Momany} Y.,  {Moretti} A.,  {Milone} A.~P.,    {Ortolani} S.,  2009, A\&A,
  493, 959

\bibitem[\protect\citeauthoryear{{Brown} \& {Wallerstein}}{{Brown} \&
  {Wallerstein}}{1993}]{Brown1993}
{Brown} J.~A.,  {Wallerstein} G.,  1993, AJ, 106, 133

\bibitem[\protect\citeauthoryear{{Calamida} et~al.,}{{Calamida}
  et~al.}{2005}]{Calamida2005}
{Calamida} A.,  et~al., 2005, ApJL, 634, L69

\bibitem[\protect\citeauthoryear{{Campbell}, {Yong}, {Wylie-de Boer},
  {Stancliffe}, {Lattanzio}, {Angelou}, {Grundahl} \& {Sneden}}{{Campbell}
  et~al.}{2010}]{Campbell2010}
{Campbell} S.~W.,  {Yong} D.,  {Wylie-de Boer} E.~C.,  {Stancliffe} R.~J.,
  {Lattanzio} J.~C.,  {Angelou} G.~C.,  {Grundahl} F.,    {Sneden} C.,  2010,
  Memorie della Societa Astronomica Italiana, 81, 1004

\bibitem[\protect\citeauthoryear{{Carretta}, {Gratton}, {Lucatello},
  {Bragaglia} \& {Bonifacio}}{{Carretta} et~al.}{2005}]{Carretta2005}
{Carretta} E.,  {Gratton} R.~G.,  {Lucatello} S.,  {Bragaglia} A.,
  {Bonifacio} P.,  2005, A\&A, 433, 597

\bibitem[\protect\citeauthoryear{{Castelli} \& {Kurucz}}{{Castelli} \&
  {Kurucz}}{2004}]{Castelli2004}
{Castelli} F.,  {Kurucz} R.~L.,  2004, ArXiv Astrophysics e-prints

\bibitem[\protect\citeauthoryear{{D'Antona}, {D'Ercole}, {Marino}, {Milone},
  {Ventura} \& {Vesperini}}{{D'Antona} et~al.}{2011}]{D'Antona2011}
{D'Antona} F.,  {D'Ercole} A.,  {Marino} A.~F.,  {Milone} A.~P.,  {Ventura} P.,
     {Vesperini} E.,  2011, ApJ, 736, 5

\bibitem[\protect\citeauthoryear{{D'Cruz}, {O'Connell}, {Rood}, {Whitney},
  {Dorman}, {Landsman}, {Hill}, {Stecher} \& {Bohlin}}{{D'Cruz}
  et~al.}{2000}]{DCruz2000}
{D'Cruz} N.~L.,  {O'Connell} R.~W.,  {Rood} R.~T.,  {Whitney} J.~H.,  {Dorman}
  B.,  {Landsman} W.~B.,  {Hill} R.~S.,  {Stecher} T.~P.,    {Bohlin} R.~C.,
  2000, ApJ, 530, 352

\bibitem[\protect\citeauthoryear{{Decin}, {Shkedy}, {Molenberghs}, {Aerts} \&
  {Aerts}}{{Decin} et~al.}{2004}]{Decin2004}
{Decin} L.,  {Shkedy} Z.,  {Molenberghs} G.,  {Aerts} M.,    {Aerts} C.,  2004,
  A\&A, 421, 281

\bibitem[\protect\citeauthoryear{{D'Ercole}, {D'Antona}, {Ventura}, {Vesperini}
  \& {McMillan}}{{D'Ercole} et~al.}{2010}]{DErcole2010}
{D'Ercole} A.,  {D'Antona} F.,  {Ventura} P.,  {Vesperini} E.,    {McMillan}
  S.~L.~W.,  2010, MNRAS, 407, 854

\bibitem[\protect\citeauthoryear{{D'Ercole}, {Vesperini}, {D'Antona},
  {McMillan} \& {Recchi}}{{D'Ercole} et~al.}{2008}]{DErcole2008}
{D'Ercole} A.,  {Vesperini} E.,  {D'Antona} F.,  {McMillan} S.~L.~W.,
  {Recchi} S.,  2008, MNRAS, 391, 825

\bibitem[\protect\citeauthoryear{{Dupree}, {Strader} \& {Smith}}{{Dupree}
  et~al.}{2011}]{Dupree2011}
{Dupree} A.~K.,  {Strader} J.,    {Smith} G.~H.,  2011, ApJ, 728, 155

\bibitem[\protect\citeauthoryear{{Gratton}, {Johnson}, {Lucatello}, {D'Orazi}
  \& {Pilachowski}}{{Gratton} et~al.}{2011}]{Gratton2011}
{Gratton} R.~G.,  {Johnson} C.~I.,  {Lucatello} S.,  {D'Orazi} V.,
  {Pilachowski} C.,  2011, A\&A, 534, A72

\bibitem[\protect\citeauthoryear{{Harbeck}, {Smith} \& {Grebel}}{{Harbeck}
  et~al.}{2003}]{Harbeck2003}
{Harbeck} D.,  {Smith} G.~H.,    {Grebel} E.~K.,  2003, AJ, 125, 197

\bibitem[\protect\citeauthoryear{Hartigan \& Wong}{Hartigan \&
  Wong}{1979}]{Hartigan1979}
Hartigan J.~A.,  Wong M.~A.,  1979, Journal of the Royal Statistical Society.
  Series C (Applied Statistics), 28, 100

\bibitem[\protect\citeauthoryear{{Herwig}, {VandenBerg}, {Navarro}, {Ferguson}
  \& {Paxton}}{{Herwig} et~al.}{2012}]{Herwig2012}
{Herwig} F.,  {VandenBerg} D.~A.,  {Navarro} J.~F.,  {Ferguson} J.,    {Paxton}
  B.,  2012, ArXiv e-prints

\bibitem[\protect\citeauthoryear{{Hinkle}, {Wallace}, {Valenti} \&
  {Harmer}}{{Hinkle} et~al.}{2000}]{Hinkle2000}
{Hinkle} K.,  {Wallace} L.,  {Valenti} J.,    {Harmer} D.,  2000, {Visible and
  Near Infrared Atlas of the Arcturus Spectrum 3727-9300 A}.
Astronomical Society of the Pacific

\bibitem[\protect\citeauthoryear{{Johnson}, {Kraft}, {Pilachowski}, {Sneden},
  {Ivans} \& {Benman}}{{Johnson} et~al.}{2005}]{Johnson2005}
{Johnson} C.~I.,  {Kraft} R.~P.,  {Pilachowski} C.~A.,  {Sneden} C.,  {Ivans}
  I.~I.,    {Benman} G.,  2005, PASP, 117, 1308

\bibitem[\protect\citeauthoryear{{Johnson} \& {Pilachowski}}{{Johnson} \&
  {Pilachowski}}{2010}]{Johnson2010}
{Johnson} C.~I.,  {Pilachowski} C.~A.,  2010, ApJ, 722, 1373

\bibitem[\protect\citeauthoryear{{Johnson}, {Pilachowski}, {Simmerer} \&
  {Schwenk}}{{Johnson} et~al.}{2008}]{Johnson2008}
{Johnson} C.~I.,  {Pilachowski} C.~A.,  {Simmerer} J.,    {Schwenk} D.,  2008,
  ApJ, 681, 1505

\bibitem[\protect\citeauthoryear{{King}, {Bedin}, {Cassisi}, {Milone},
  {Bellini}, {Piotto}, {Anderson}, {Pietrinferni} \& {Cordier}}{{King}
  et~al.}{2012}]{King2012}
{King} I.~R.,  {Bedin} L.~R.,  {Cassisi} S.,  {Milone} A.~P.,  {Bellini} A.,
  {Piotto} G.,  {Anderson} J.,  {Pietrinferni} A.,    {Cordier} D.,  2012, AJ,
  144, 5

\bibitem[\protect\citeauthoryear{{Marino}, {Milone}, {Piotto}, {Cassisi},
  {D'Antona}, {Anderson}, {Aparicio}, {Bedin}, {Renzini} \&
  {Villanova}}{{Marino} et~al.}{2012}]{Marino2012}
{Marino} A.~F.,  {Milone} A.~P.,  {Piotto} G.,  {Cassisi} S.,  {D'Antona} F.,
  {Anderson} J.,  {Aparicio} A.,  {Bedin} L.~R.,  {Renzini} A.,    {Villanova}
  S.,  2012, ApJ, 746, 14

\bibitem[\protect\citeauthoryear{{Marino}, {Milone}, {Piotto}, {Villanova},
  {Gratton}, {D'Antona}, {Anderson}, {Bedin}, {Bellini}, {Cassisi}, {Geisler},
  {Renzini} \& {Zoccali}}{{Marino} et~al.}{2011}]{Marino2011}
{Marino} A.~F.,  {Milone} A.~P.,  {Piotto} G.,  {Villanova} S.,  {Gratton} R.,
  {D'Antona} F.,  {Anderson} J.,  {Bedin} L.~R.,  {Bellini} A.,  {Cassisi} S.,
  {Geisler} D.,  {Renzini} A.,    {Zoccali} M.,  2011, ApJ, 731, 64

\bibitem[\protect\citeauthoryear{Norris}{Norris}{2012}]{Norris2012}
Norris J.,  2012, CN, CH region line list, private communication

\bibitem[\protect\citeauthoryear{{Norris}, {Cottrell}, {Freeman} \& {Da
  Costa}}{{Norris} et~al.}{1981}]{Norris1981}
{Norris} J.,  {Cottrell} P.~L.,  {Freeman} K.~C.,    {Da Costa} G.~S.,  1981,
  ApJ, 244, 205

\bibitem[\protect\citeauthoryear{{Norris} \& {Da Costa}}{{Norris} \& {Da
  Costa}}{1995}]{Norris1995}
{Norris} J.~E.,  {Da Costa} G.~S.,  1995, ApJ, 447, 680

\bibitem[\protect\citeauthoryear{{Pancino}, {Ferraro}, {Bellazzini}, {Piotto}
  \& {Zoccali}}{{Pancino} et~al.}{2000}]{Pancino2000}
{Pancino} E.,  {Ferraro} F.~R.,  {Bellazzini} M.,  {Piotto} G.,    {Zoccali}
  M.,  2000, ApJL, 534, L83

\bibitem[\protect\citeauthoryear{{Pietrinferni}, {Cassisi}, {Salaris} \&
  {Castelli}}{{Pietrinferni} et~al.}{2006}]{Pietrinferni2006}
{Pietrinferni} A.,  {Cassisi} S.,  {Salaris} M.,    {Castelli} F.,  2006, ApJ,
  642, 797

\bibitem[\protect\citeauthoryear{{Piotto}, {Villanova}, {Bedin}, {Gratton},
  {Cassisi}, {Momany}, {Recio-Blanco}, {Lucatello}, {Anderson}, {King},
  {Pietrinferni} \& {Carraro}}{{Piotto} et~al.}{2005}]{Piotto2005}
{Piotto} G.,  {Villanova} S.,  {Bedin} L.~R.,  {Gratton} R.,  {Cassisi} S.,
  {Momany} Y.,  {Recio-Blanco} A.,  {Lucatello} S.,  {Anderson} J.,  {King}
  I.~R.,  {Pietrinferni} A.,    {Carraro} G.,  2005, ApJ, 621, 777

\bibitem[\protect\citeauthoryear{{R Development Core Team}}{{R Development Core
  Team}}{2011}]{R2011}
{R Development Core Team} 2011, R: A Language and Environment for Statistical
  Computing.
R Foundation for Statistical Computing, Vienna, Austria

\bibitem[\protect\citeauthoryear{{Rey}, {Lee}, {Ree}, {Joo}, {Sohn} \&
  {Walker}}{{Rey} et~al.}{2004}]{Rey2004}
{Rey} S.-C.,  {Lee} Y.-W.,  {Ree} C.~H.,  {Joo} J.-M.,  {Sohn} Y.-J.,
  {Walker} A.~R.,  2004, AJ, 127, 958

\bibitem[\protect\citeauthoryear{{Rutten}}{{Rutten}}{1978}]{Rutten1978}
{Rutten} R.~J.,  1978, Solar Physics, 56, 237

\bibitem[\protect\citeauthoryear{{Sbordone}, {Salaris}, {Weiss} \&
  {Cassisi}}{{Sbordone} et~al.}{2011}]{Sbordone2011}
{Sbordone} L.,  {Salaris} M.,  {Weiss} A.,    {Cassisi} S.,  2011, A\&A, 534,
  A9

\bibitem[\protect\citeauthoryear{{Skrutskie} et~al.,}{{Skrutskie}
  et~al.}{2006}]{Skrutskie2006}
{Skrutskie} M.~F.,  et~al., 2006, AJ, 131, 1163

\bibitem[\protect\citeauthoryear{{Sneden}}{{Sneden}}{1973}]{Sneden1973}
{Sneden} C.~A.,  1973, PhD thesis, The University of Texas at Austin

\bibitem[\protect\citeauthoryear{{Sobeck}, {Kraft}, {Sneden}, {Preston},
  {Cowan}, {Smith}, {Thompson}, {Shectman} \& {Burley}}{{Sobeck}
  et~al.}{2011}]{Sobeck2011}
{Sobeck} J.~S.,  {Kraft} R.~P.,  {Sneden} C.,  {Preston} G.~W.,  {Cowan} J.~J.,
   {Smith} G.~H.,  {Thompson} I.~B.,  {Shectman} S.~A.,    {Burley} G.~S.,
  2011, AJ, 141, 175

\bibitem[\protect\citeauthoryear{{Sollima}, {Pancino}, {Ferraro}, {Bellazzini},
  {Straniero} \& {Pasquini}}{{Sollima} et~al.}{2005}]{Sollima2005}
{Sollima} A.,  {Pancino} E.,  {Ferraro} F.~R.,  {Bellazzini} M.,  {Straniero}
  O.,    {Pasquini} L.,  2005, ApJ, 634, 332

\bibitem[\protect\citeauthoryear{{Stanford}, {Da Costa} \& {Norris}}{{Stanford}
  et~al.}{2010}]{Stanford2010}
{Stanford} L.~M.,  {Da Costa} G.~S.,    {Norris} J.~E.,  2010, ApJ, 714, 1001

\bibitem[\protect\citeauthoryear{{Stanford}, {Da Costa}, {Norris} \&
  {Cannon}}{{Stanford} et~al.}{2006}]{Stanford2006}
{Stanford} L.~M.,  {Da Costa} G.~S.,  {Norris} J.~E.,    {Cannon} R.~D.,  2006,
  ApJ, 647, 1075

\bibitem[\protect\citeauthoryear{{Taylor}}{{Taylor}}{2005}]{Taylor2005}
{Taylor} M.~B.,  2005, in {P.~Shopbell, M.~Britton, \& R.~Ebert} ed.,
  Astronomical Data Analysis Software and Systems XIV Vol.~347 of Astronomical
  Society of the Pacific Conference Series, {TOPCAT \& STIL: Starlink
  Table/VOTable Processing Software}.
pp 29--+

\bibitem[\protect\citeauthoryear{{van Leeuwen}, {Le Poole}, {Reijns}, {Freeman}
  \& {de Zeeuw}}{{van Leeuwen} et~al.}{2000}]{vanLeeuwen2000}
{van Leeuwen} F.,  {Le Poole} R.~S.,  {Reijns} R.~A.,  {Freeman} K.~C.,    {de
  Zeeuw} P.~T.,  2000, A\&A, 360, 472

\bibitem[\protect\citeauthoryear{{van Loon}, {van Leeuwen}, {Smalley}, {Smith},
  {Lyons}, {McDonald} \& {Boyer}}{{van Loon} et~al.}{2007}]{vanLoon2007}
{van Loon} J.~T.,  {van Leeuwen} F.,  {Smalley} B.,  {Smith} A.~W.,  {Lyons}
  N.~A.,  {McDonald} I.,    {Boyer} M.~L.,  2007, MNRAS, 382, 1353

\bibitem[\protect\citeauthoryear{{Villanova}, {Carraro}, {Scarpa} \&
  {Marconi}}{{Villanova} et~al.}{2010}]{Villanova2010}
{Villanova} S.,  {Carraro} G.,  {Scarpa} R.,    {Marconi} G.,  2010, New
  Astronomy, 15, 520

\bibitem[\protect\citeauthoryear{{Villanova}, {Piotto}, {King}, {Anderson},
  {Bedin}, {Gratton}, {Cassisi}, {Momany}, {Bellini}, {Cool}, {Recio-Blanco} \&
  {Renzini}}{{Villanova} et~al.}{2007}]{Villanova2007}
{Villanova} S.,  {Piotto} G.,  {King} I.~R.,  {Anderson} J.,  {Bedin} L.~R.,
  {Gratton} R.~G.,  {Cassisi} S.,  {Momany} Y.,  {Bellini} A.,  {Cool} A.~M.,
  {Recio-Blanco} A.,    {Renzini} A.,  2007, ApJ, 663, 296

\bibitem[\protect\citeauthoryear{{Worley} \& {Cottrell}}{{Worley} \&
  {Cottrell}}{2012}]{Worley2012}
{Worley} C.~C.,  {Cottrell} P.~L.,  2012, PASA, 29, 29

\end{thebibliography}

\bsp

\label{lastpage}

\end{document}